\documentclass{aa}

\usepackage{graphicx}
\usepackage{amsmath}	
\usepackage{booktabs}
\usepackage{newtxtext,newtxmath}
\usepackage{xcolor}
\usepackage{pdflscape}
\usepackage{hyperref}

\begin{document}

   \title{Gas excitation of post-starburst galaxies at $0.6 < z < 1.3$}
   
   \author{A. Zanella
          \inst{1}
          \and
          S. Belli
          \inst{2}
          \and
          F. M. Valentino
          \inst{3,4}
          \and
          A. Bolamperti
          \inst{5,6}
          }

   \institute{Istituto Nazionale di Astrofisica (INAF), Via Gobetti 93/3, 40129 Bologna, Italy;
              \email{anita.zanella@inaf.it}
         \and
    Dipartimento di Fisica e Astronomia, Università di Bologna, Italy
         \and
        Cosmic Dawn Center (DAWN), Denmark
        \and
        DTU Space, Technical University of Denmark, Elektrovej 327, DK-2800, Kgs. Lyngby, Denmark
        \and
        Max-Planck-Institut für Astrophysik, Karl-Schwarzschild-Str. 1, D-85748 Garching, Germany
        \and
        INAF -- IASF Milano, via A. Corti 12, I-20133 Milano, Italy
             }

   \date{Received XXX; accepted XXX}
 
 \abstract
   {Molecular gas in galaxies traces both the fuel for star formation and the processes that enhance or suppress it. Observing its physical state (e.g., excitation) can reveal when and why galaxies stop forming stars.}
   {We observed the CO(5-4) emission of 8 post-starburst galaxies at $z \sim 0.6 - 1.3$. To our knowledge, this is the first time that high-$J$ transitions are probed for post-starburst or quiescent galaxies beyond the local Universe. All of them have detections in lower-$J$ CO transitions (either CO(2-1) or CO(3-2)) and molecular gas fractions up to $\sim 20\%$. By studying the ratio $R_{52} = L'\mathrm{CO(5-4)}/L'\mathrm{CO(2-1)}$, a proxy for the gas excitation, we aim to constrain the physical state of the gas.}
   {The CO excitation helps distinguishing among different mechanisms responsible for the low star formation efficiency
(SFE) of post-starburst galaxies. In the first scenario, the molecular gas is predominantly diffuse and cold, implying a low fraction of dense star-forming gas and in turn low $R_{52}$ values. In the second scenario, elevated gas temperatures at moderate densities, for example due to AGN activity, shocks, or enhanced turbulence, would instead produce high $R_{52}$ values. 
}
   {Our post-starbursts have on average $R_{52} = 0.28$, comparable to high-redshift main-sequence galaxies. However, when considering only the CO(5-4) non-detections, which also coincide with post-starbusts that do not show signs of interaction, we obtain $R_{52} < 0.10$, twice lower than local star-forming galaxies and more than 2.5 times lower than high-redshift sources. The average CO Spectral Line Energy Distribution (SLED) peaks at $J = 3$, similar to the Milky Way. Three galaxies show signs of interactions (tidal features, companions). They have $R_{52} = 0.40$ and SLEDs peaking at $J \gtrsim 4-5$. In at least one case additional mechanisms (e.g., AGN, shocks) are needed to explain the steep rise of the SLED up to $J = 5$.}
   {Our results favor a scenario in which most systems are dominated by low-density molecular gas with low excitation, consistent with quenching driven by gas stabilization, feedback regulation, or stripping. In interacting systems instead, enhanced excitation is likely driven by heating processes not related to star-formation (e.g., AGN, turbulence, shocks). Residual star formation is insufficient to rapidly exhaust the remaining molecular gas in the majority of post-starburst galaxies.}


   \maketitle

\section{Introduction}

Despite significant observational and theoretical efforts, the physical mechanisms quenching star formation remain under active investigation and are still a central challenge in our understanding of galaxy evolution. A plethora of different mechanisms have been proposed to stop star formation by either reducing or depleting the molecular gas reservoir or rendering the gas unable to cool and form stars \citep{Man2018}. In particular, cold gas accretion onto the galaxy dark matter halo might be prevented \citep{Larson1980, Feldmann2015, Peng2015, Feldmann2017, Dave2017} or the infalling gas might be heated up because of virial shocks \citep{Rees1977, Dekel2006}. In both cases,  additional mechanisms are needed to prevent the cold gas reservoir that is already present in the galaxy to form new stars and to keep the galaxy quiescent. Feedback-driven outflows from active galactic nuclei (AGN) might expel the gas from the galaxy (quasar-mode feedback, \citealt{DiMatteo2005}) while inefficient accretion onto the supermassive back hole (radio-mode feedback) keeps the galaxy quenched by heating the gas \citep{Bower2006, Croton2006, Hopkins2006}. Other mechanisms that have been suggested are gravitational heating due to the energy released by ram-pressure stripping \citep{Gunn1972, Werle2022} or dynamical friction (e.g., during the infall of a satellite or clump, \citealt{Khochfar2008}). Gas might also be consumed by starbursts (SBs) possibly triggered by major mergers \citep{Mihos1996, Springel2005, Wellons2015} and disk instabilities \citep{Zolotov2015}. However, quenching might be due to a low star formation efficiency (SFE) rather than due to the lack of fuel. The cold gas disk of galaxies might be stable against collapse and fragmentation due to the presence of a bulge \citep{Martig2009, Johansson2009} or a stellar bar \citep{Khoperskov2018}.

Understanding what mechanisms are responsible for galaxy quenching requires identifying galaxies in the immediate aftermath of their final star-forming episode and accurately characterizing their cold gas content and physical conditions. Post-starburst (post-SB) galaxies, also referred to as "k+a" or "e+a" systems, represent a promising population for probing this transition phase. These galaxies exhibit optical spectra dominated by strong Balmer absorption lines, indicative of A-type stars formed in a recent burst of star formation, coupled with weak or absent nebular emission lines, signaling little to no ongoing star formation \citep{Dressler1983, Couch1987}.
Recent observations of low-$J$ CO transitions such as CO(2-1) and CO(3-2), classical tracers of total cold molecular gas, have revealed significant molecular gas reservoirs in post-SB galaxies up to redshifts of $z \sim 1.4$ \citep{French2015, Rowlands2015, Suess2017, French2018, Belli2021, Bezanson2021, Zanella2023, Suess2025}. These studies report a wide variety of gas fractions that can be as high as 20\%, suggesting that substantial amounts of star-forming fuel can remain in galaxies after the cessation of star formation. Despite this, post-SBs appear to lie systematically below the star-forming galaxy population in the Kennicutt–Schmidt relation, implying that their low star formation rates are not due to a lack of gas, but rather to a suppression of SFE \citep{Zanella2023}.
This interpretation, however, remains tentative due to two key uncertainties: the excitation correction factor required to convert CO line luminosities to total molecular gas masses, typically assumed to be 15 – 50\%, based on values from star-forming galaxies; and the obscured star formation rate (SFR), which is often inferred from mid-infrared (e.g., 24$\mu$m) emission. The latter can be significantly contaminated by dust heating from evolved stellar populations, particularly in post-starburst systems \citep{Leja2019, Belli2021}, complicating the accurate measurement of residual star formation activity. 

Probing denser and/or warmer molecular gas through CO(5–4) emission provides a key diagnostic of the excitation state of the molecular medium, commonly quantified via the ratios $R_{52}, R_{53}$ (i.e., $L'_\mathrm{CO54}/L'_\mathrm{CO21}$ and $L'_\mathrm{CO54}/L'_\mathrm{CO32}$, respectively). These ratios are sensitive to both gas density and kinetic temperature and can be used to assess if residual, possibly obscured star formation is present, or whether the molecular gas is predominantly diffuse and inefficient at forming stars. If large reservoirs of molecular gas traced by low-$J$ CO transitions are unable to sustain star formation in post-starburst galaxies, low $R_{52}$ and $R_{53}$ values are expected, indicating a small fraction of dense and/or warm gas relative to the cold, diffuse molecular component. Conversely, high excitation ratios comparable to those of main-sequence or starburst galaxies may signal either the presence of dense, star-forming gas (implying a high star formation efficiency) or elevated gas temperatures driven by additional heating mechanisms such as merger-induced shocks, turbulence, or AGN activity. 
To date, however, observations of high-$J$ CO transitions in post-starburst or quenched galaxies remain scarce.

In the local Universe, \cite{French2023} constrained the CO Spectral Line Energy Distribution (SLED) of a sample of post-starburst galaxies up to the CO(3–2) transition. They found low excitation, consistent with the low HCO$^+$/CO ratios observed in the same galaxies, indicating that low fractions of dense molecular gas relative to the total molecular gas drive the reduced star formation efficiency (SFE) in these systems. This sample partially overlaps with that of \cite{Smercina2021}, who analyzed high-resolution CO(2–1) observations and found that the turbulent kinetic pressure of the gas exceeds that of typical star-forming disks, resembling instead the conditions in local interacting galaxies such as the Antennae. They suggest that this turbulence prevents the gas from collapsing, further contributing to the suppression of star formation. Constraining the CO excitation at higher redshift is therefore key to understand whether similar mechanisms drive the low SFE of post-starbursts across redshift.

This paper is organized as follows: in Sec. \ref{sec:data} we describe the
sample and the data; in Sec. \ref{sec:analysis} we discuss how we estimated the observables and derived physical properties; in Sec. \ref{sec:results} we present our results also in the context of available literature; in Sec. \ref{sec:discussion} we discuss the possible mechanisms that keep our galaxies quiescent and the decoupling of molecular gas and dust; in Sec. \ref{sec:conclusions} we conclude and summarize our findings. Throughout the paper we adopt a flat $\Lambda$CDM cosmology with $\Omega_\mathrm{m} = 0.3$, $\Omega_\mathrm{\Lambda} = 0.7$, and $\mathrm{H_0 = 70\, km\, s^{-1}\, Mpc^{-1}}$. All magnitudes are AB magnitudes \citep{Oke1974} and we adopt a \cite{Chabrier2003} initial mass function.

\begin{table*}
    \centering
    \caption{Log of the ALMA observations used in this work.}
    \small
    \begin{tabular}{c c c c c c c c}
    \toprule
    \midrule
    ID & Date & t$_\mathrm{exp}$ & Noise R. M. S. & Beam & Program & Band & Line \\
       &      &     (min)        &   mJy/beam     & (arcsec) &   &  & \\
    (1) & (2) & (3)              &   (4)          & (5)      & (6) & (7) & (8) \\
    \midrule
    J1448 & 02 Nov 2024 & 9   & 0.082 & $1.03 \times 0.91$ & 2024.1.00061.S & 7 & CO(5-4)\\
          & 12 Mar 2018 & 100 & 0.010 & $1.31 \times 0.89$ & 2017.1.01109.S & 4 & CO(2-1)\\
          & 19 Mar 2019 & 45 & 0.023 & $1.26 \times 1.12$ & 2018.1.01264.S & 7 & CO(4-3)\\
          & 29 Mar 2019 & 150 & 0.011 & $2.22 \times 2.03$ & 2018.1.01264.S & 4 & CO(2-1)\\
    J2258 & 06 Nov 2024 & 8   & 0.083 & $1.17 \times 0.86$ & 2024.1.00061.S & 7 & CO(5-4) \\
          & 01 May 2018 & 100 & 0.009 & $2.34 \times 1.99$ & 2017.1.01109.S & 4 & CO(2-1)\\
          & 18 May 2021 & 79 & 0.008 & $0.54 \times 0.48$ & 2019.1.00221.S & 4 & CO(2-1)\\
    ID83492 & 29 Oct 2024 & 49  & 0.019 & $1.22 \times 1.15$ & 2024.1.00061.S & 6 & CO(5-4)\\
            & 09 Jan 2020 & 34 & 0.011 & $1.90 \times 1.30$ & 2019.1.00900.S & 4 & CO(3-2)\\
    J1109 & 07 Dec 2024 & 16  & 0.061 & $0.91 \times 0.80$ & 2024.1.00061.S & 7 & CO(5-4)\\
          & 10 Apr 2018 & 99 & 0.012 & $1.85 \times 1.57$ & 2017.1.01109.S & 4 & CO(2-1) \\
          & 19 Mar 2020 & 49 & 0.010 & $0.96 \times 0.70$ & 2019.1.00221.S & 4 & CO(2-1) \\
    J1302  & 31 Oct 2024 & 17  & 0.066 & $1.08 \times 0.86$ & 2024.1.00061.S & 7 & CO(5-4)\\
           & 21 Mar 2018 & 100 & 0.013 & $1.07 \times 0.98$ & 2017.1.01109.S & 4 & CO(2-1) \\
    J0912 & 25 Oct 2024 & 12  & 0.062 & $1.19 \times 0.87$ & 2024.1.00061.S & 7 & CO(5-4)\\
          & 08 Jan 2017 & 98 & 0.013 & $2.39 \times 1.91$ & 2016.1.01126.S & 4 & CO(2-1)\\
          & 24 Dec 2018 & 50 & 0.013 & $0.91 \times 0.72$ & 2018.1.01240.S & 6 & CO(4-3)\\
          & 23 Mar 2021 & 147 & 0.012 & $0.78 \times 0.71$ & 2019.1.00221.S & 4 & CO(2-1)\\
    J2202 & 21 Oct 2024 & 46  & 0.034 & $0.82 \times 0.78$ & 2024.1.00061.S & 7 & CO(5-4)\\
          & 07 Mar 2017 & 95 & 0.015 & $3.06 \times 2.16$ & 2016.1.01126.S & 4 & CO(2-1)\\
    ID97148 & 19 Nov 2024 & 110 & 0.026 & $1.39 \times 1.09$ & 2024.1.00061.S & 7 & CO(5-4) \\
            & 09 Jan 2020 & 49 & 0.010 & $1.60 \times 1.30$ & 2019.1.00900.S & 4 & CO(3-2)\\
    \bottomrule
    \end{tabular}
    \label{tab:log}    
    \tablefoot{
    Columns: (1) Galaxy ID; (2) Date of observations; (3) Integration time on source; (4) Continuum noise r. m. s.; (5) FWHM of the beam; (6) Program ID; (7) ALMA band used for the observations; (8) Targeted emission line.
    }
\end{table*}

\begin{figure*}[t!]
    \centering    
    \includegraphics[width=0.305\linewidth]{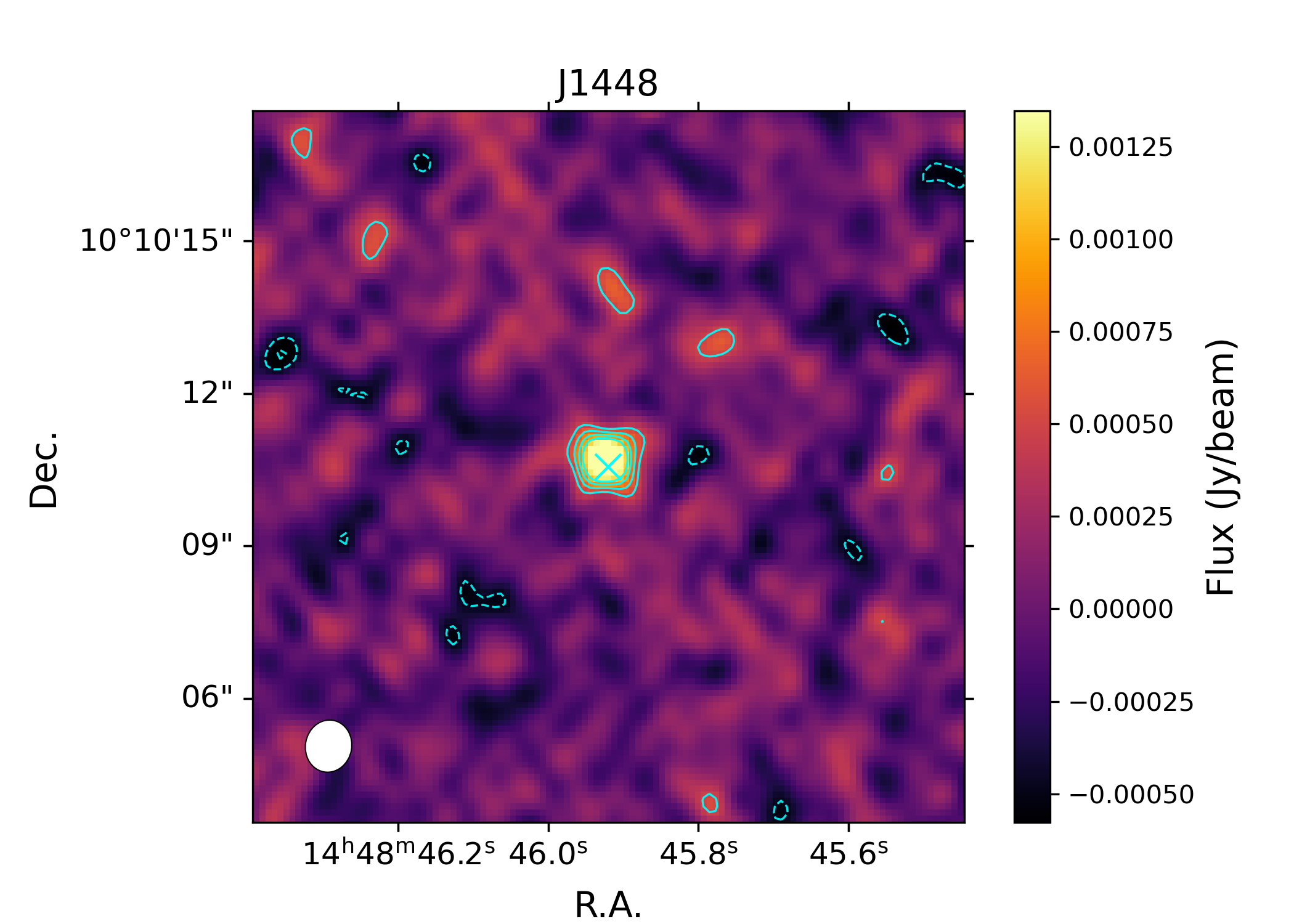}
    \includegraphics[width=0.25\linewidth]{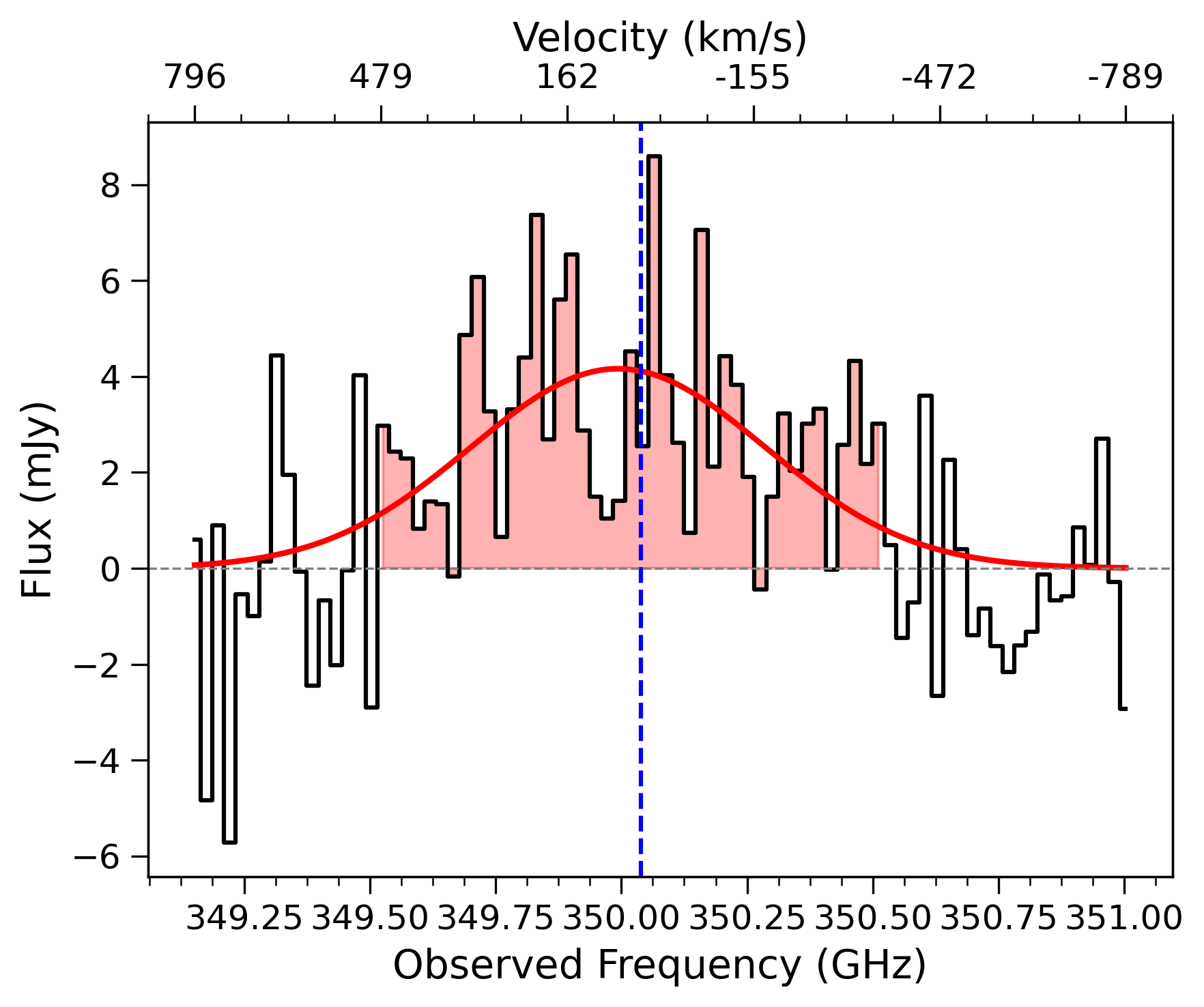}
    \includegraphics[width=0.22\linewidth]{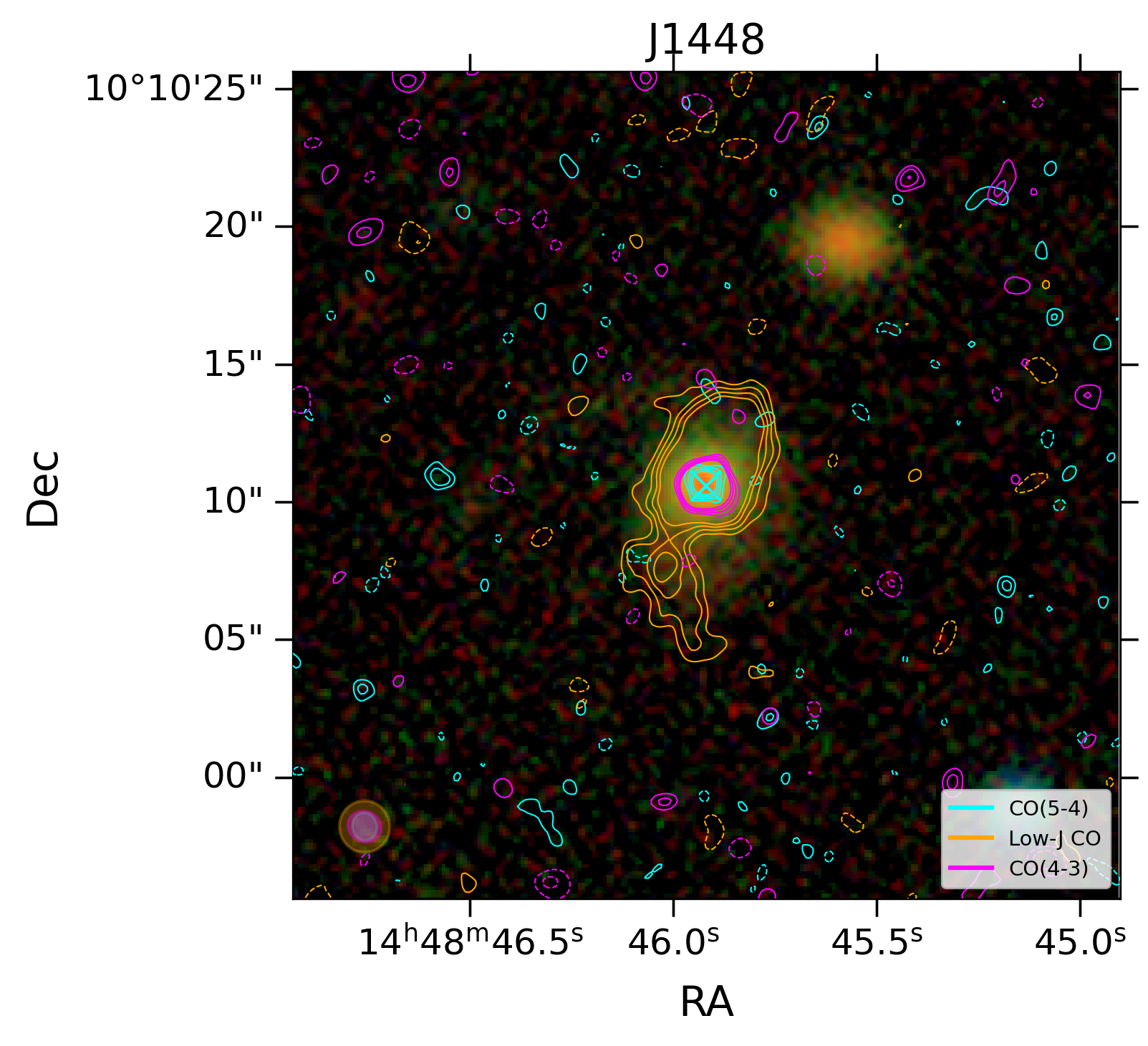}

    \includegraphics[width=0.305\linewidth]{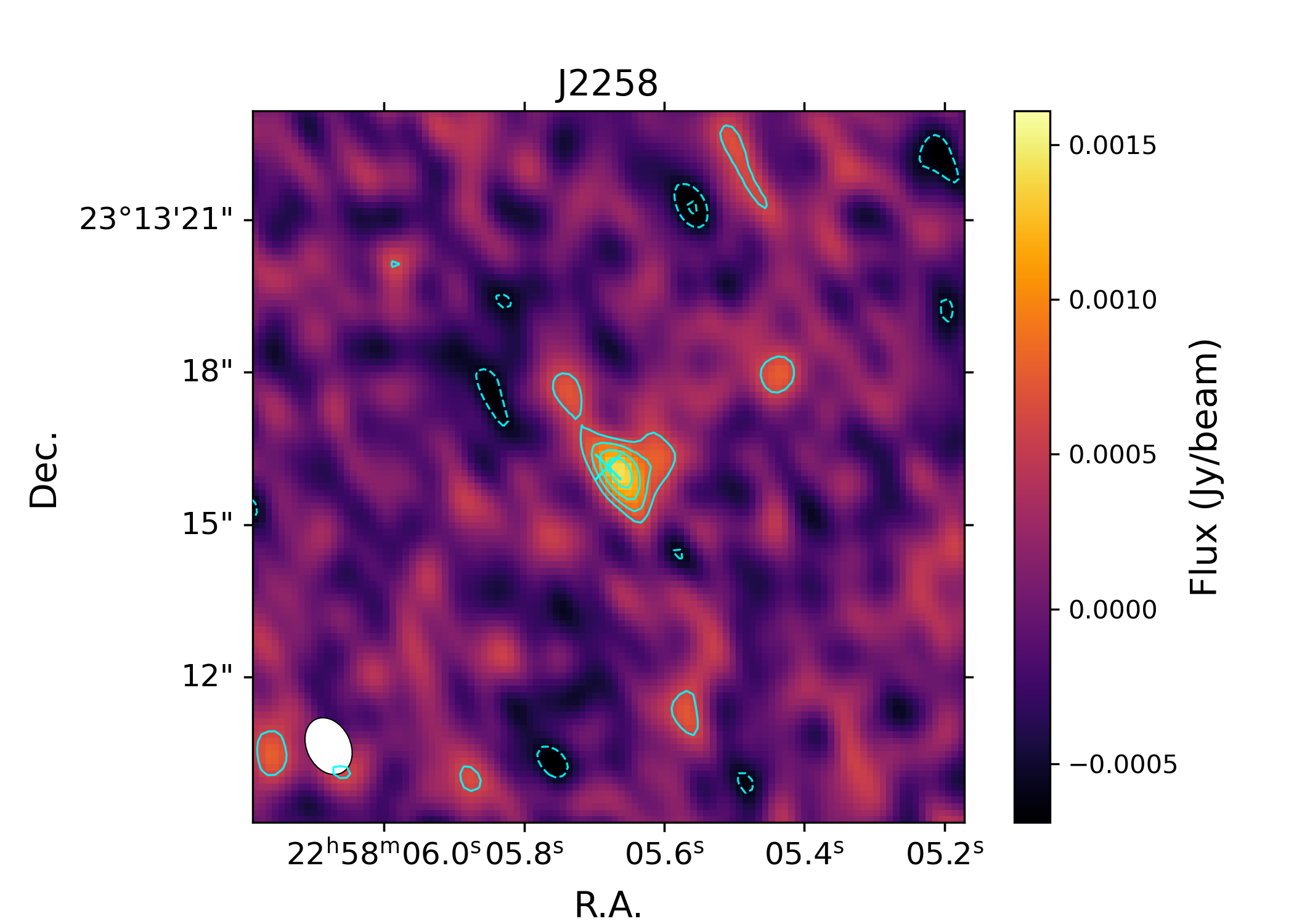}
    \includegraphics[width=0.25\linewidth]{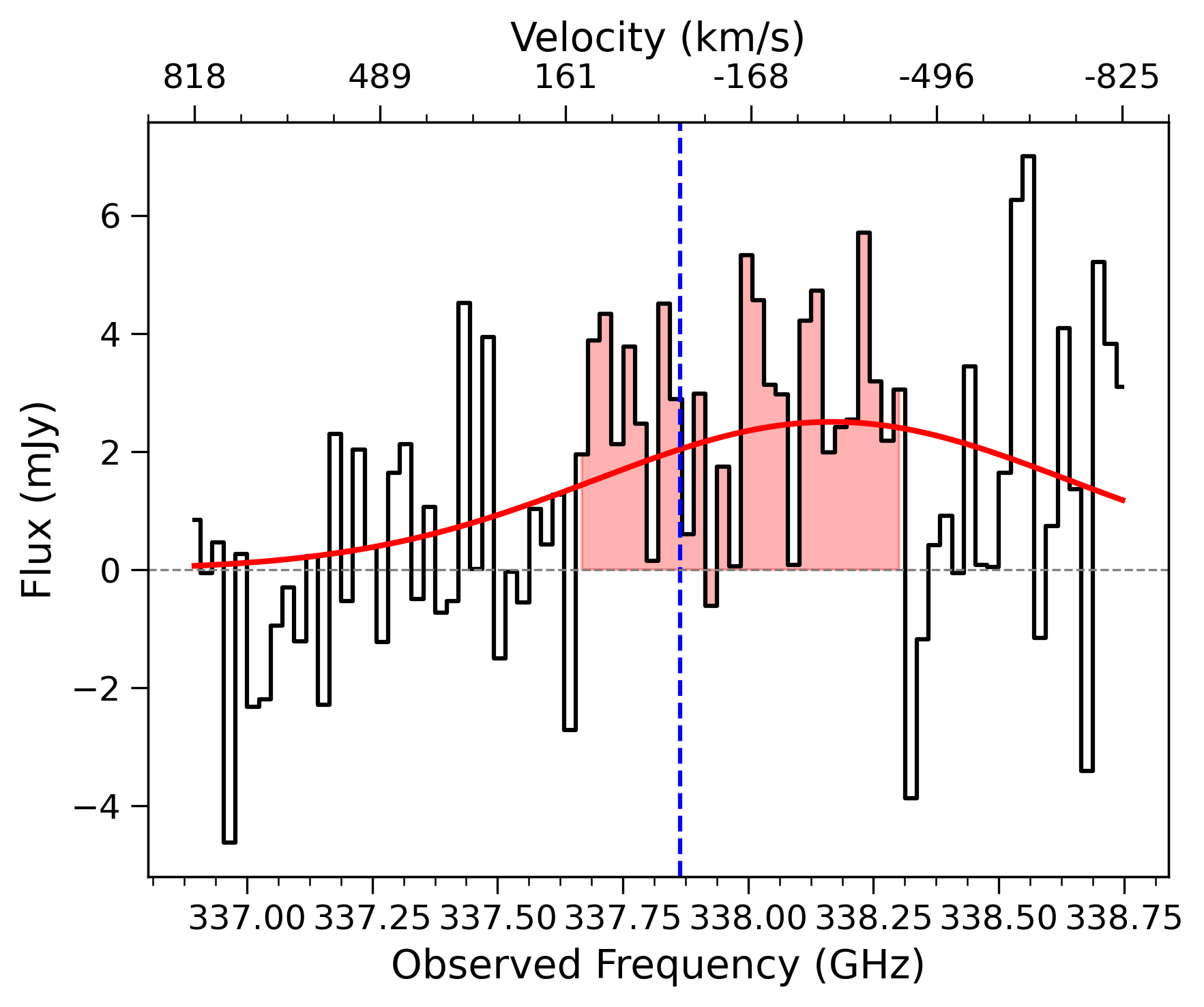}
    \includegraphics[width=0.22\linewidth]{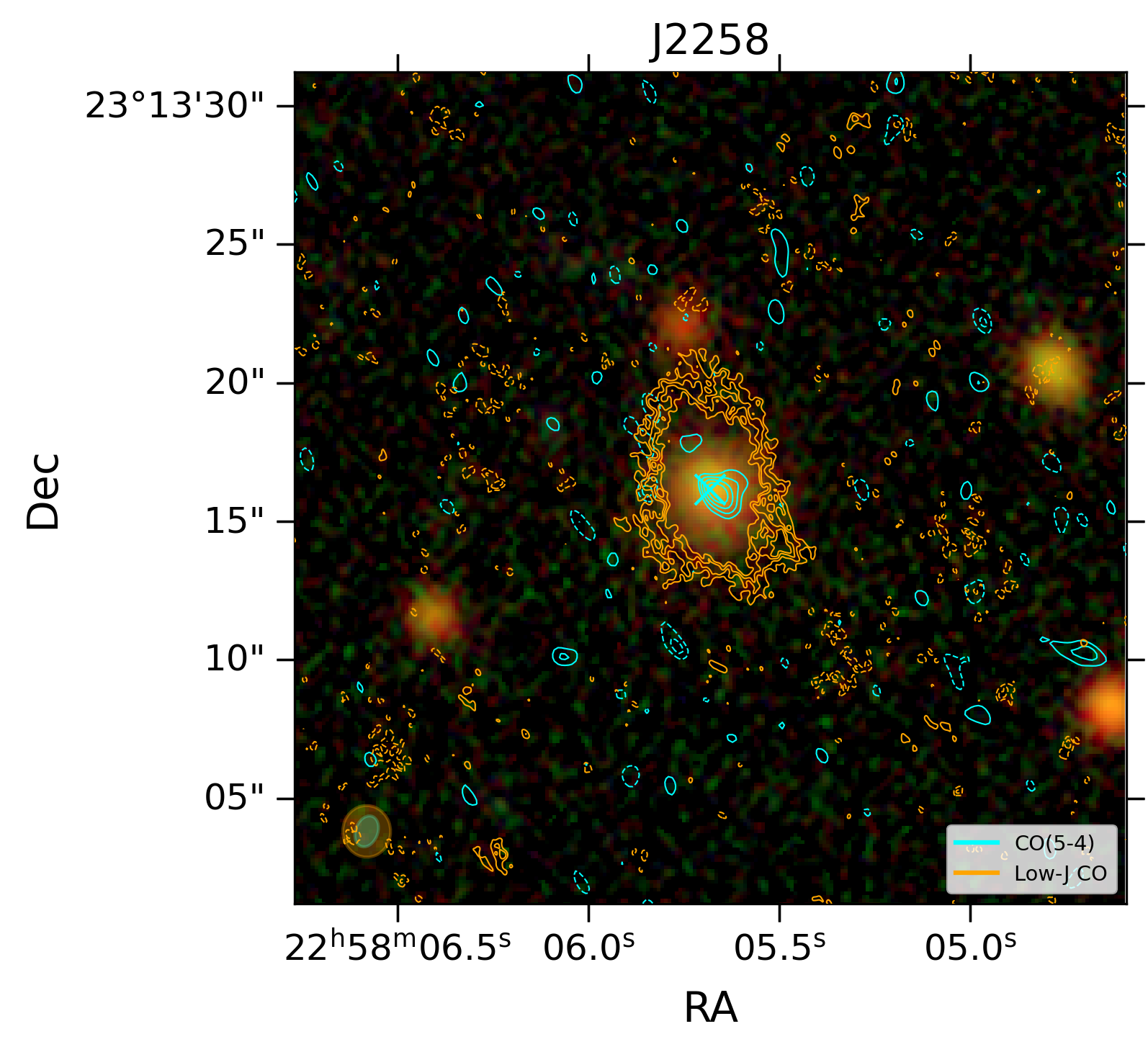}
    
    \includegraphics[width=0.305\linewidth]{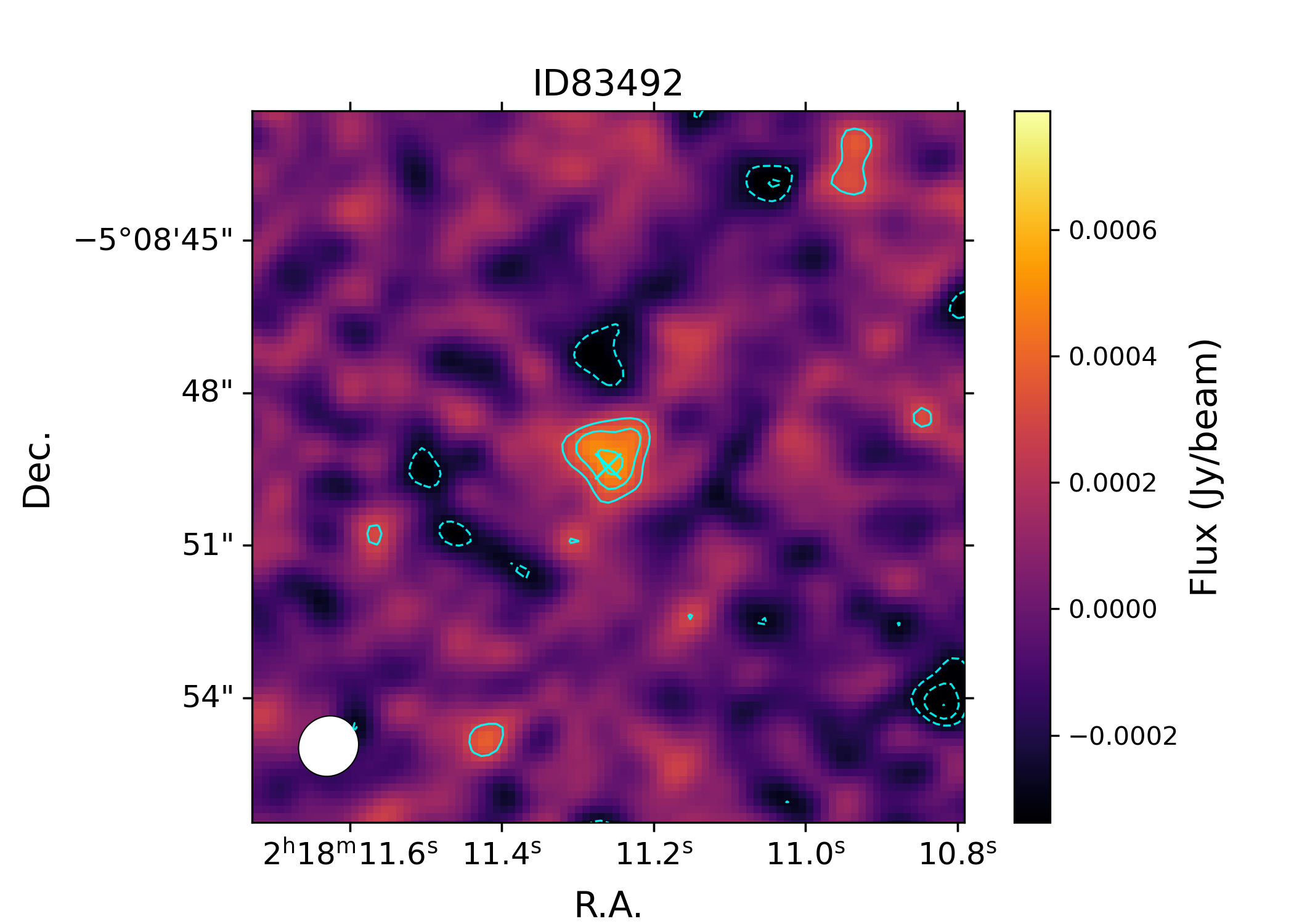}
    \includegraphics[width=0.25\linewidth]{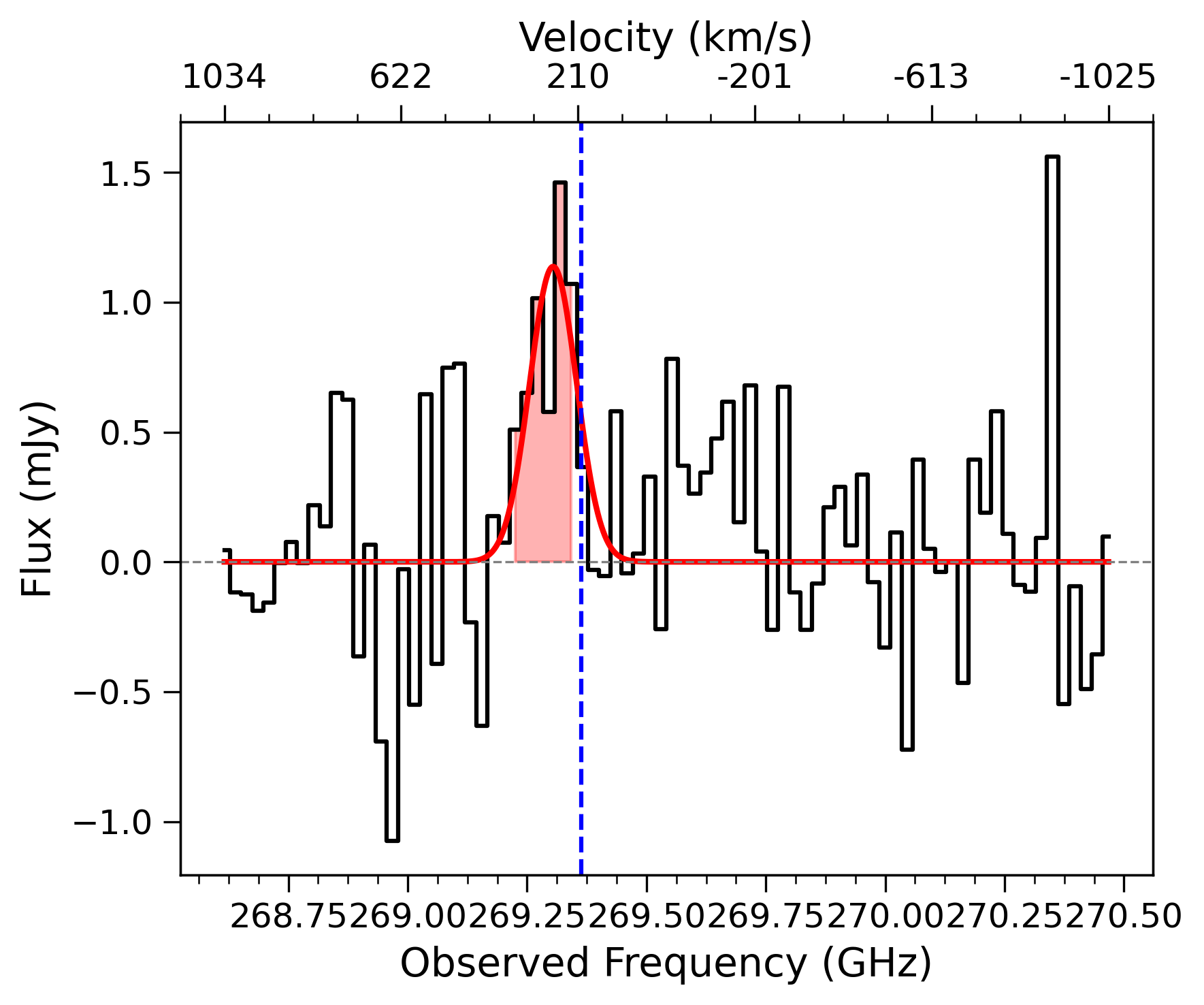}
    \includegraphics[width=0.22\linewidth]{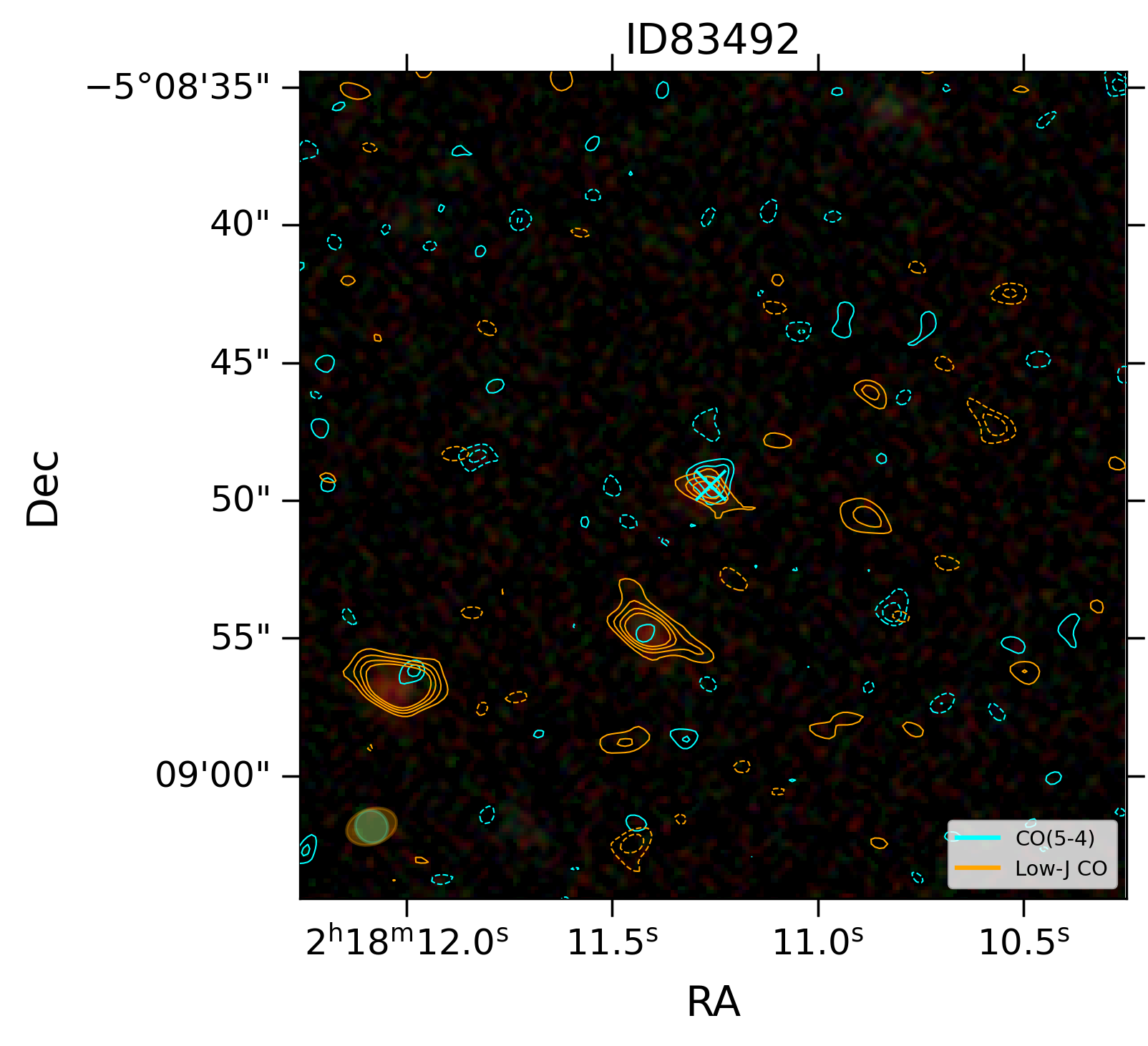}

    \includegraphics[width=0.305\linewidth]{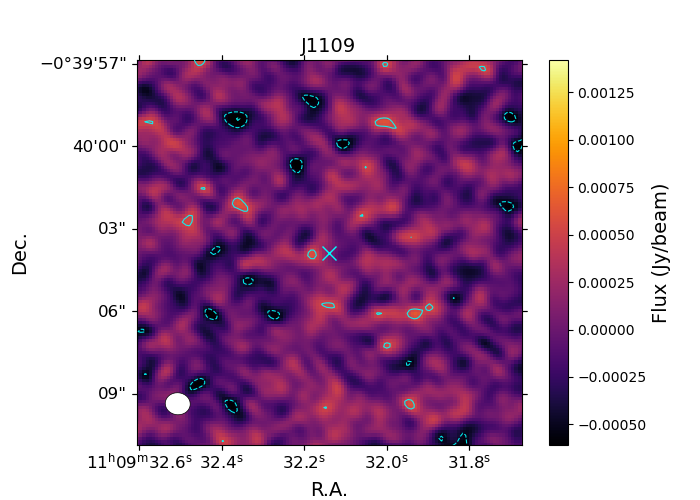}
    \includegraphics[width=0.25\linewidth]{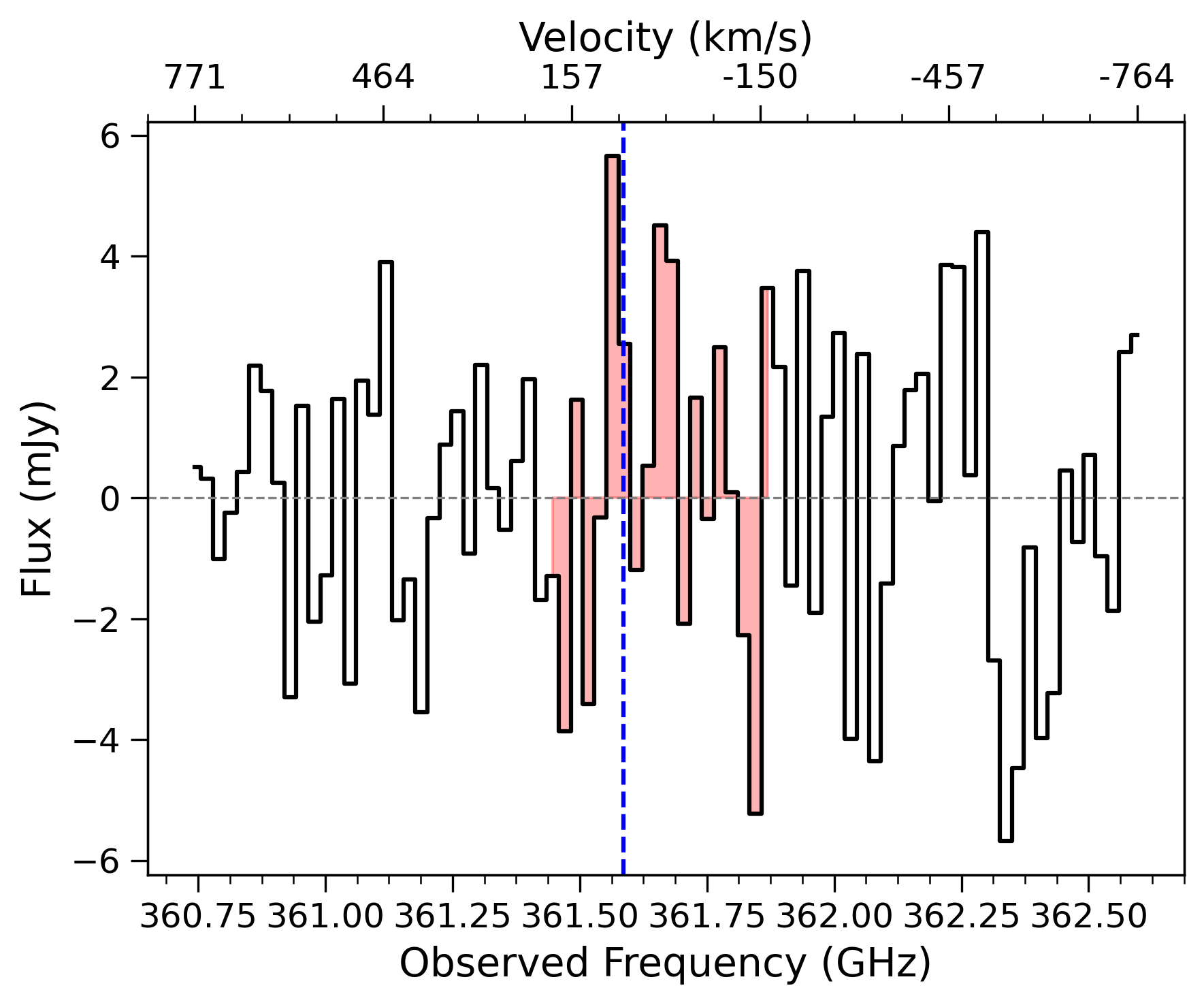}
    \includegraphics[width=0.22\linewidth]{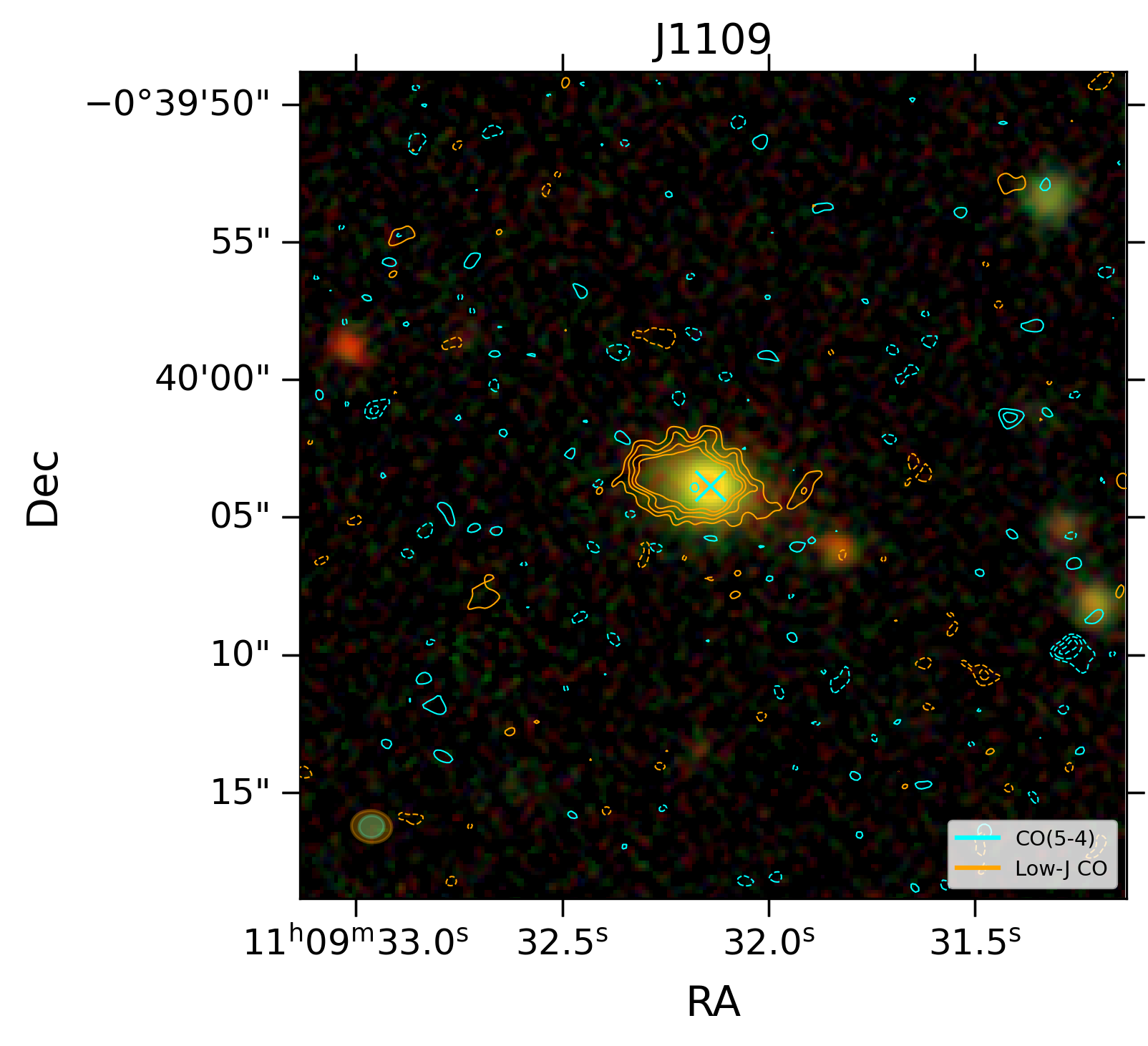}

    \caption{ALMA data of our galaxies. \textit{First column}: 2D maps of the CO(5-4) line. The cyan solid and dashed contours indicate respectively positive and negative levels of (2.5, 3.5, 4.5, 5.5) rms. The beam is reported in the bottom left corner (white ellipse). Each stamp has a size of 10" $\times$ 10". The cyan cross indicates the center of our post-SBs as estimated from the optical emission. \textit{Second column}: 1D spectra of sources extracted using a PSF to maximize the S/N. The pink shaded areas indicate the 1$\sigma$ velocity range over which we measured the CO(5-4) line flux or its upper limit. For detections we also report the Gaussian fit of the emissions (Sec. \ref{subsec:co54}). The vertical dashed lines indicate the observed frequency corresponding to the redshift of the low-$J$ CO emission. \textit{Third column}: rest-frame optical grz images from the DESI Legacy Survey DR9 \cite{Dey2019}. The cyan contours show the CO(5-4) emission, while the orange and pink indicate the low-$J$ (CO(2-1) or CO(3-2)) or CO(4-3) emission respectively. The ellipses indicate the beam size. Each stamp has a size of $20\arcsec \times 20\arcsec$. Contour levels are the same as in the first panels. (Continues on next page)}
    \label{fig:sample}
\end{figure*}

\begin{figure*}[t!]
    \centering

    \includegraphics[width=0.305\linewidth]{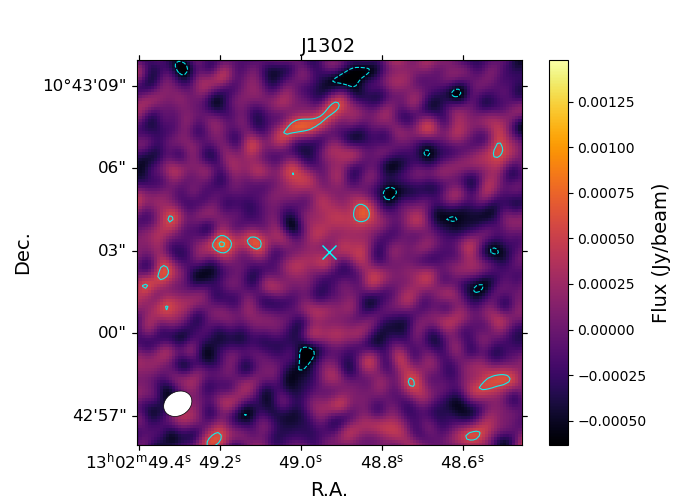}
    \includegraphics[width=0.25\linewidth]{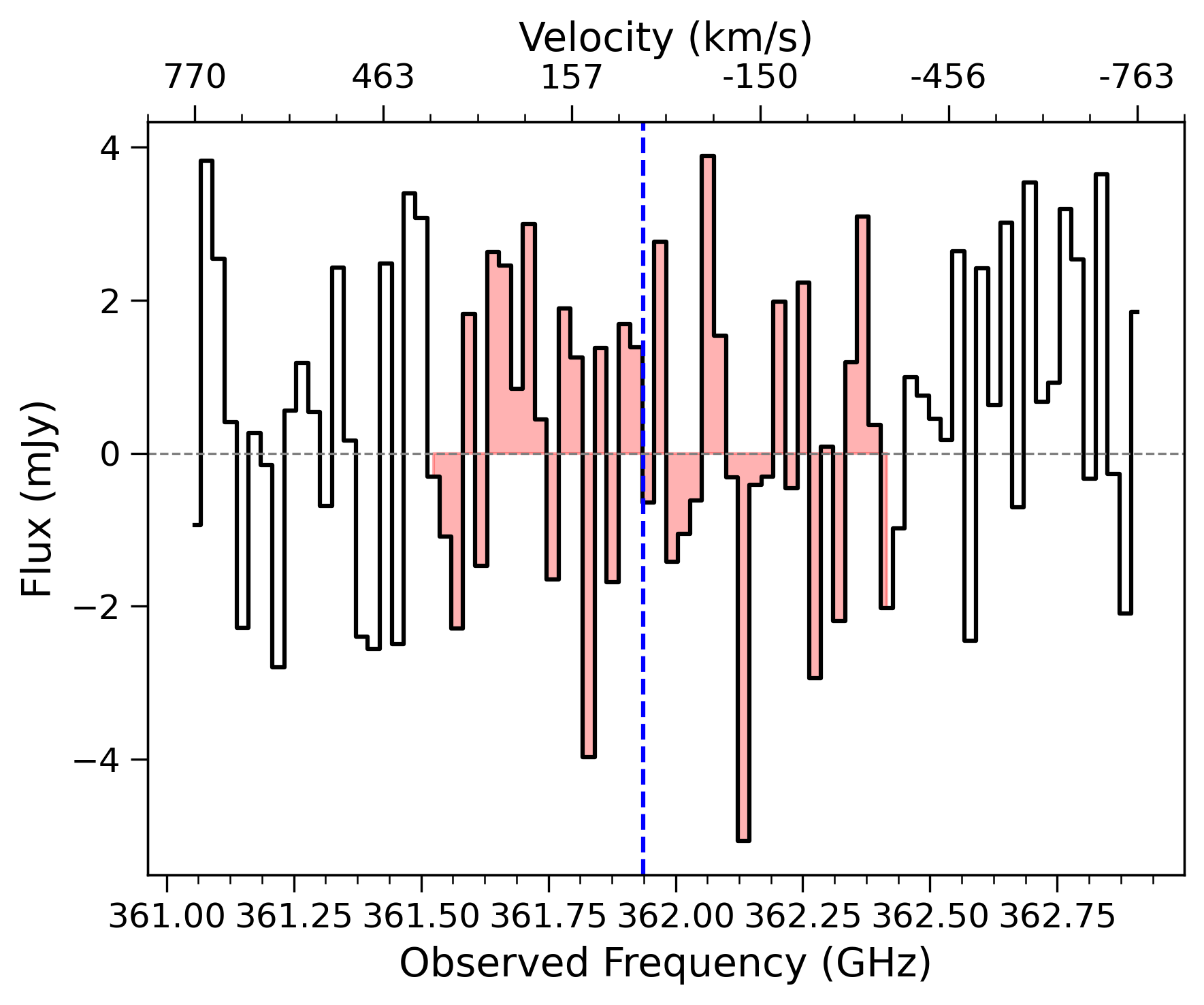}
    \includegraphics[width=0.22\linewidth]{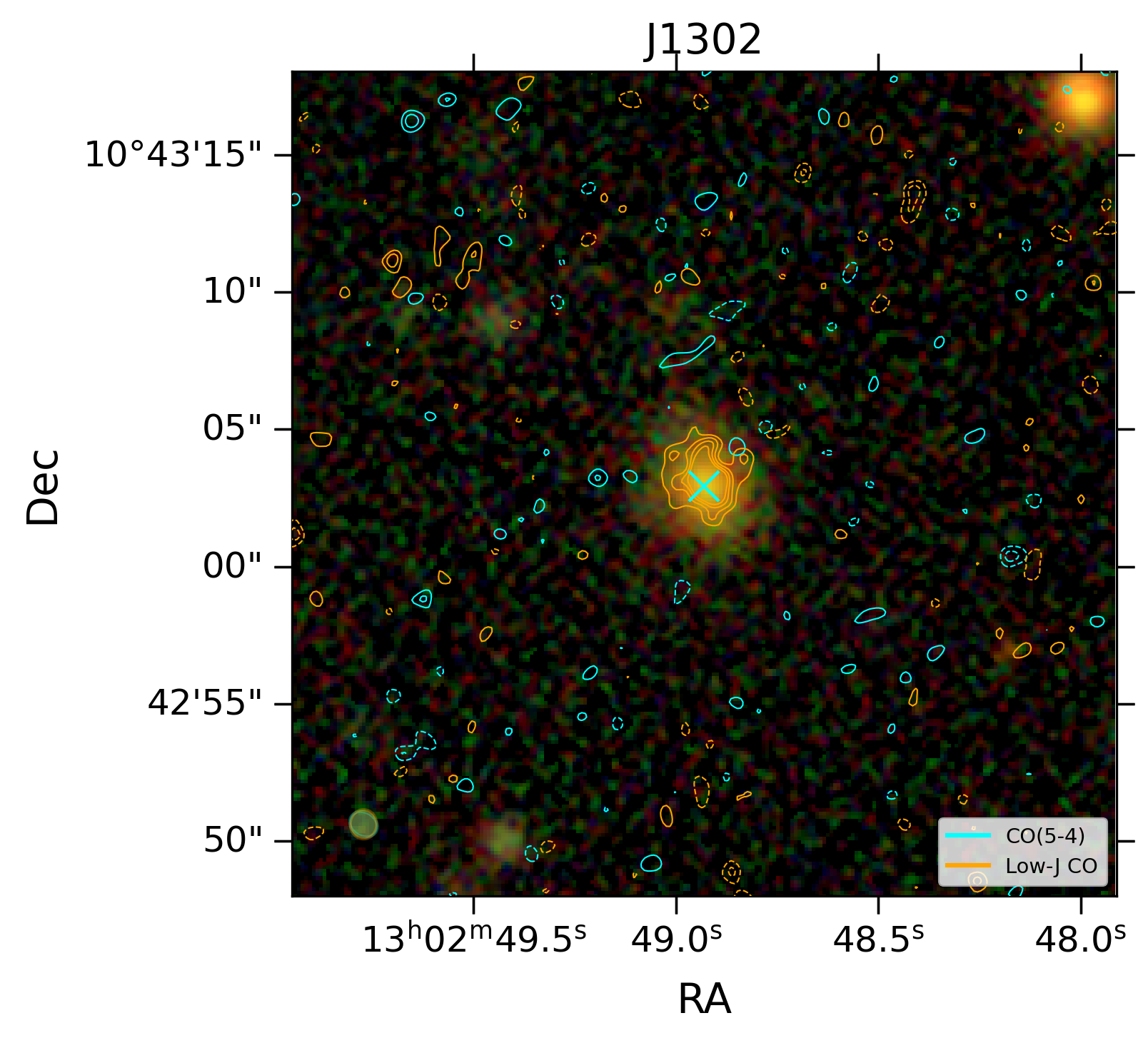}

    \includegraphics[width=0.305\linewidth]{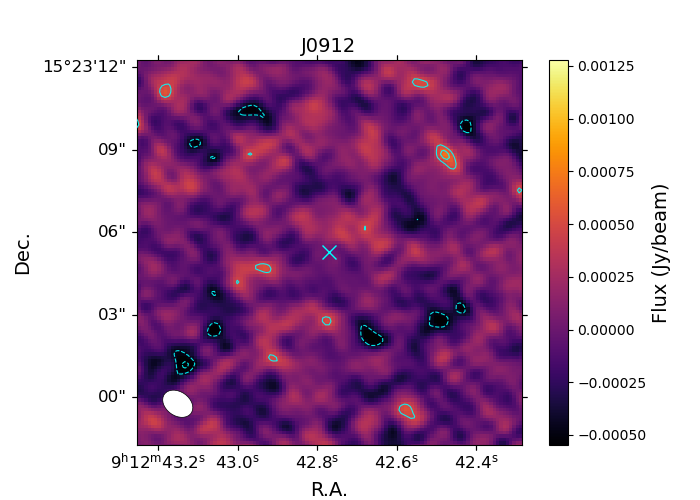}
    \includegraphics[width=0.25\linewidth]{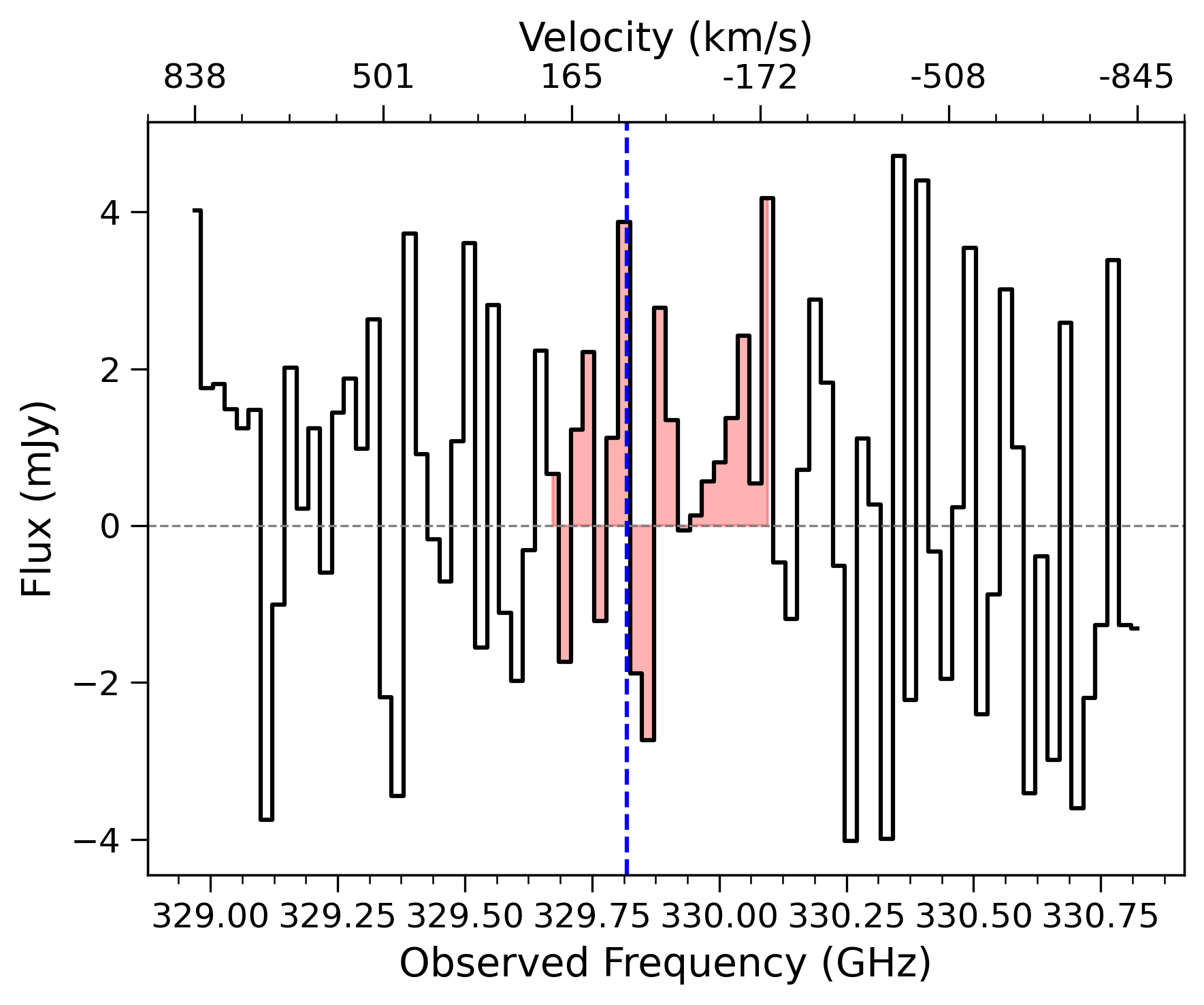}
    \includegraphics[width=0.22\linewidth]{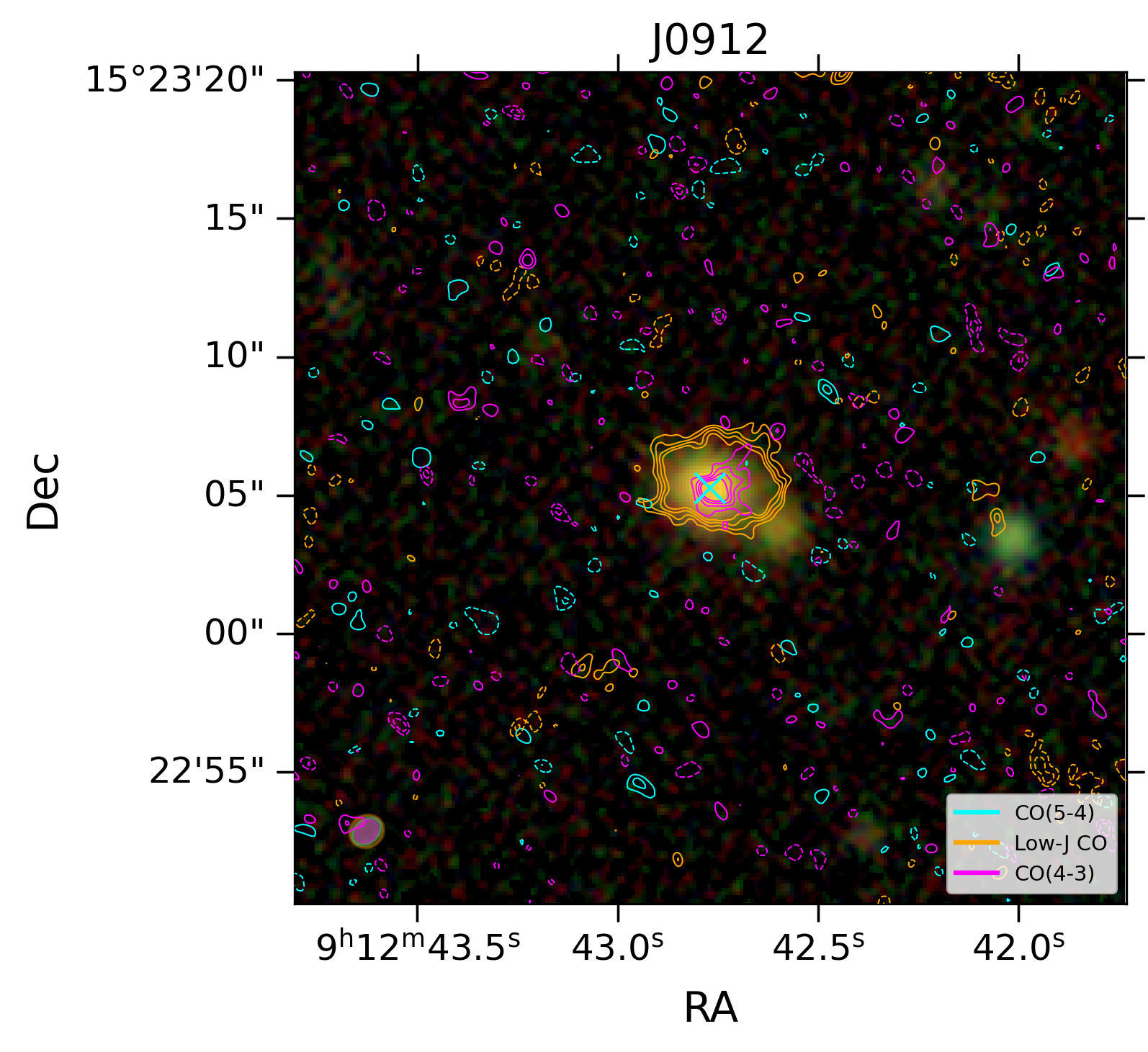}

    \includegraphics[width=0.305\linewidth]{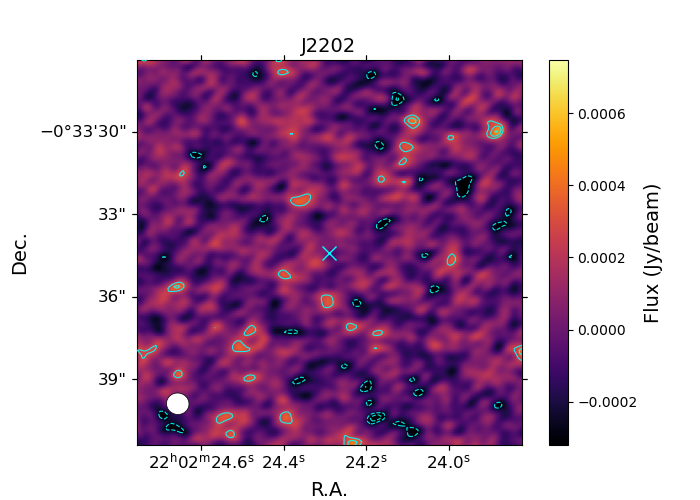}
    \includegraphics[width=0.25\linewidth]{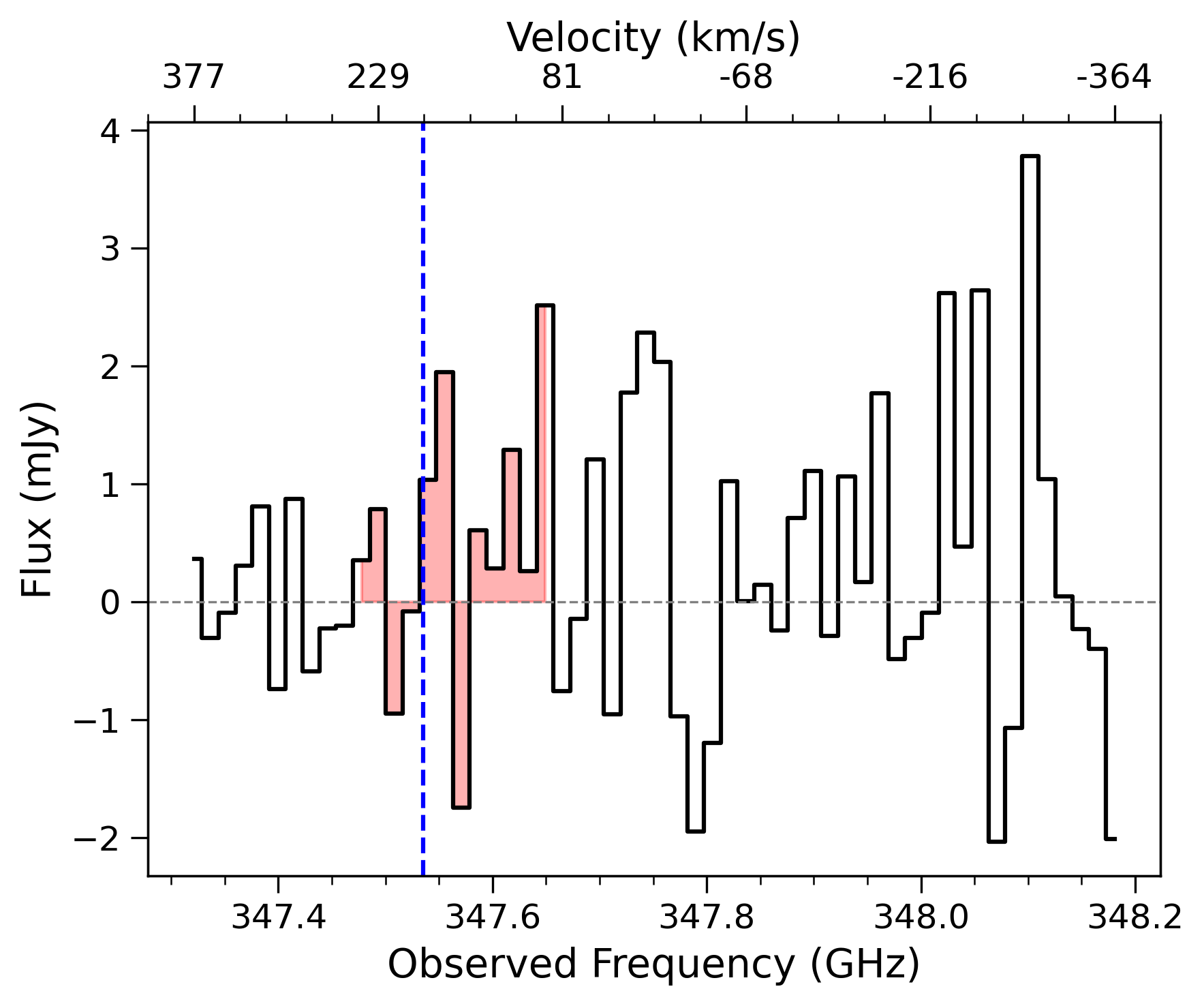}
    \includegraphics[width=0.22\linewidth]{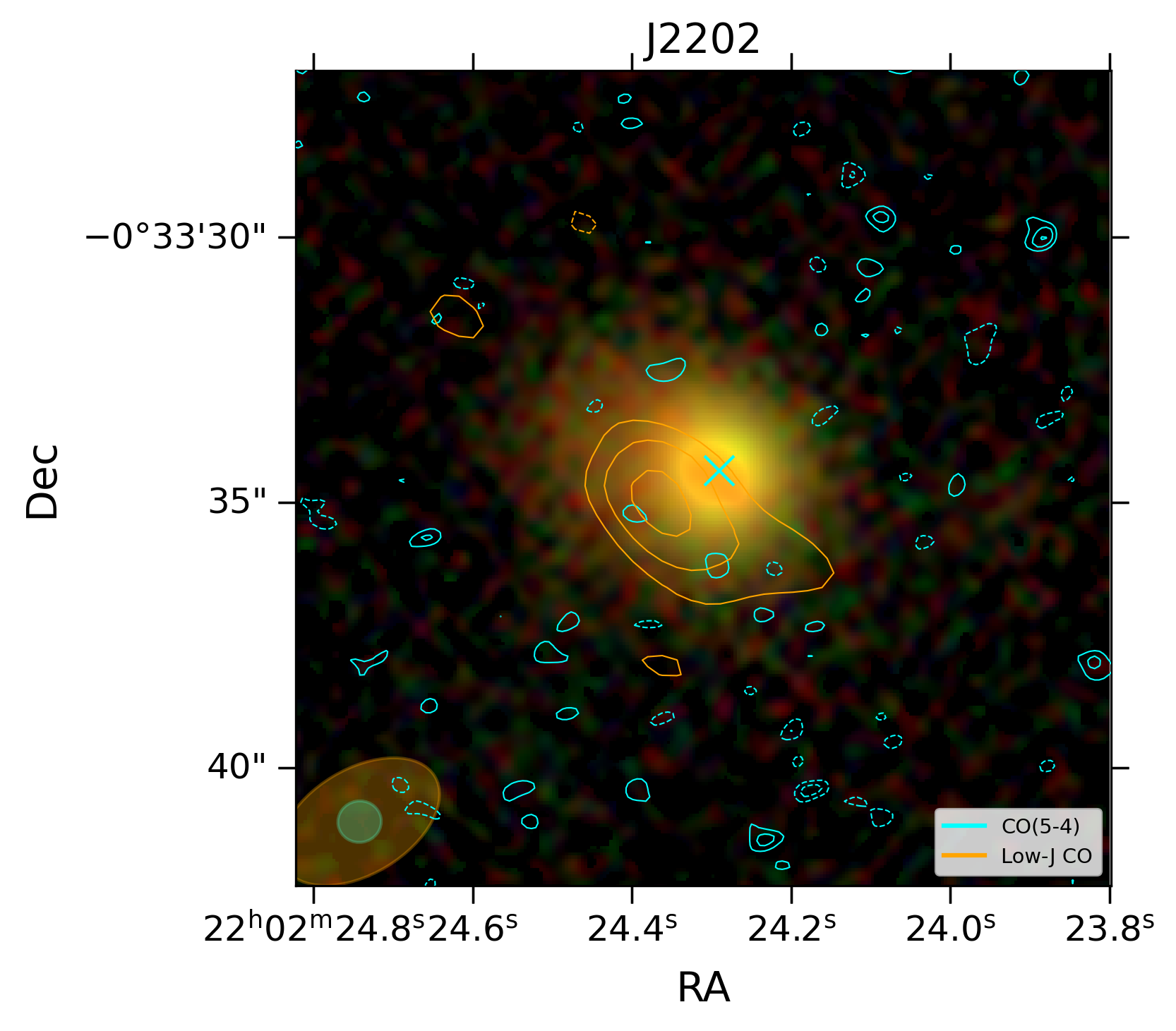}

    \includegraphics[width=0.305\linewidth]{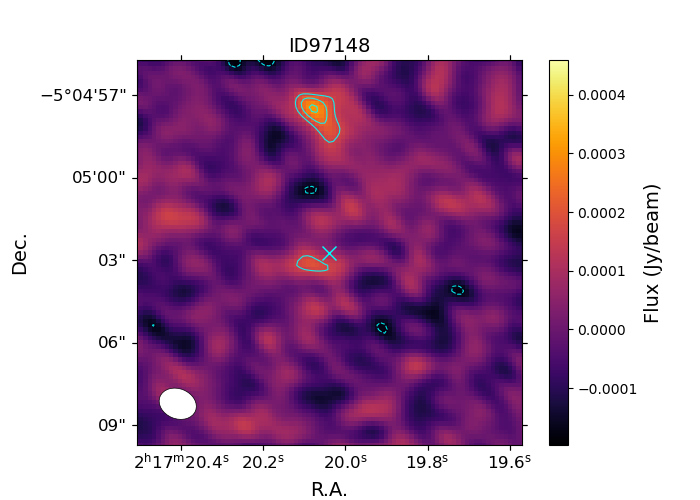}
    \includegraphics[width=0.25\linewidth]{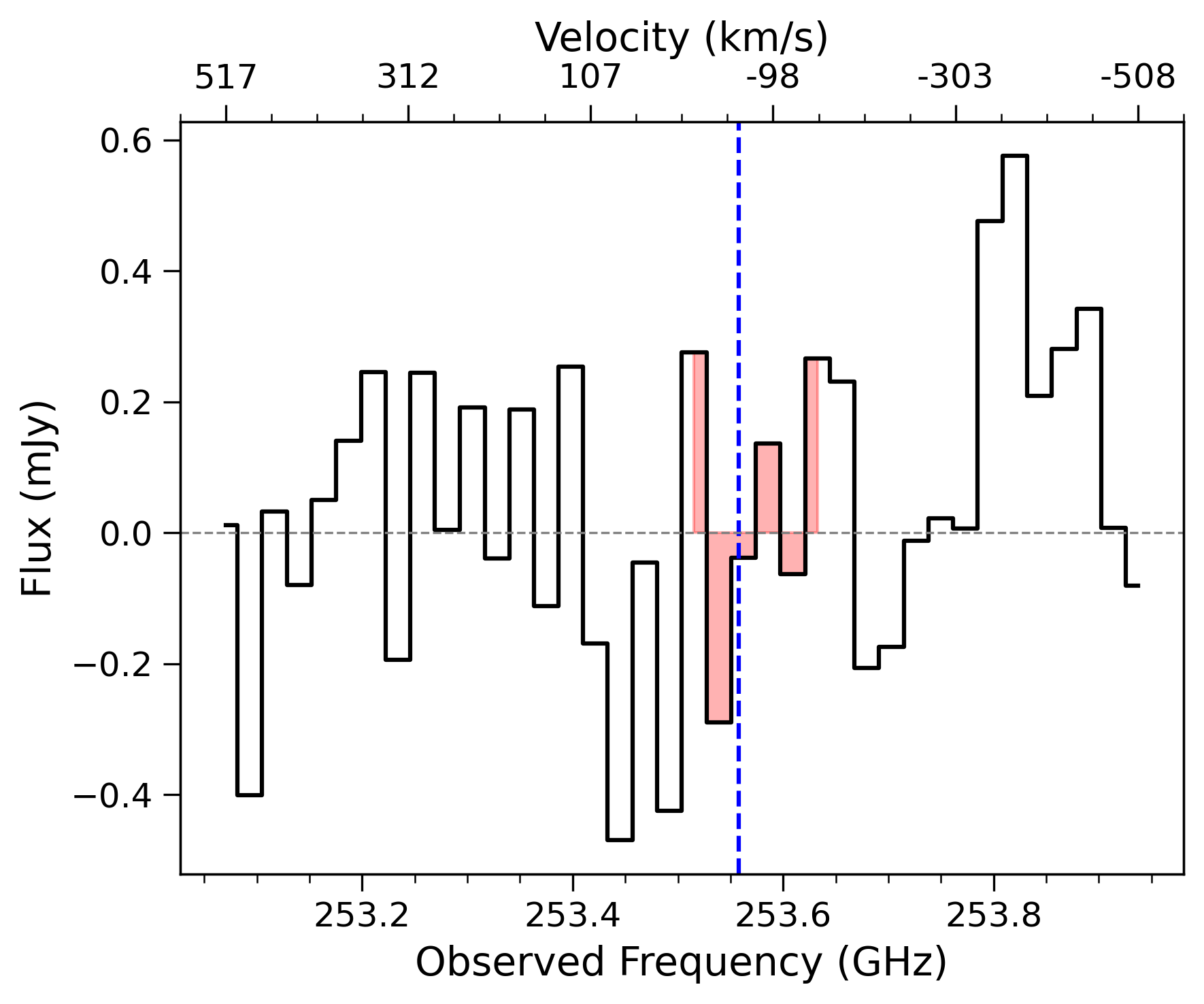}
    \includegraphics[width=0.22\linewidth]{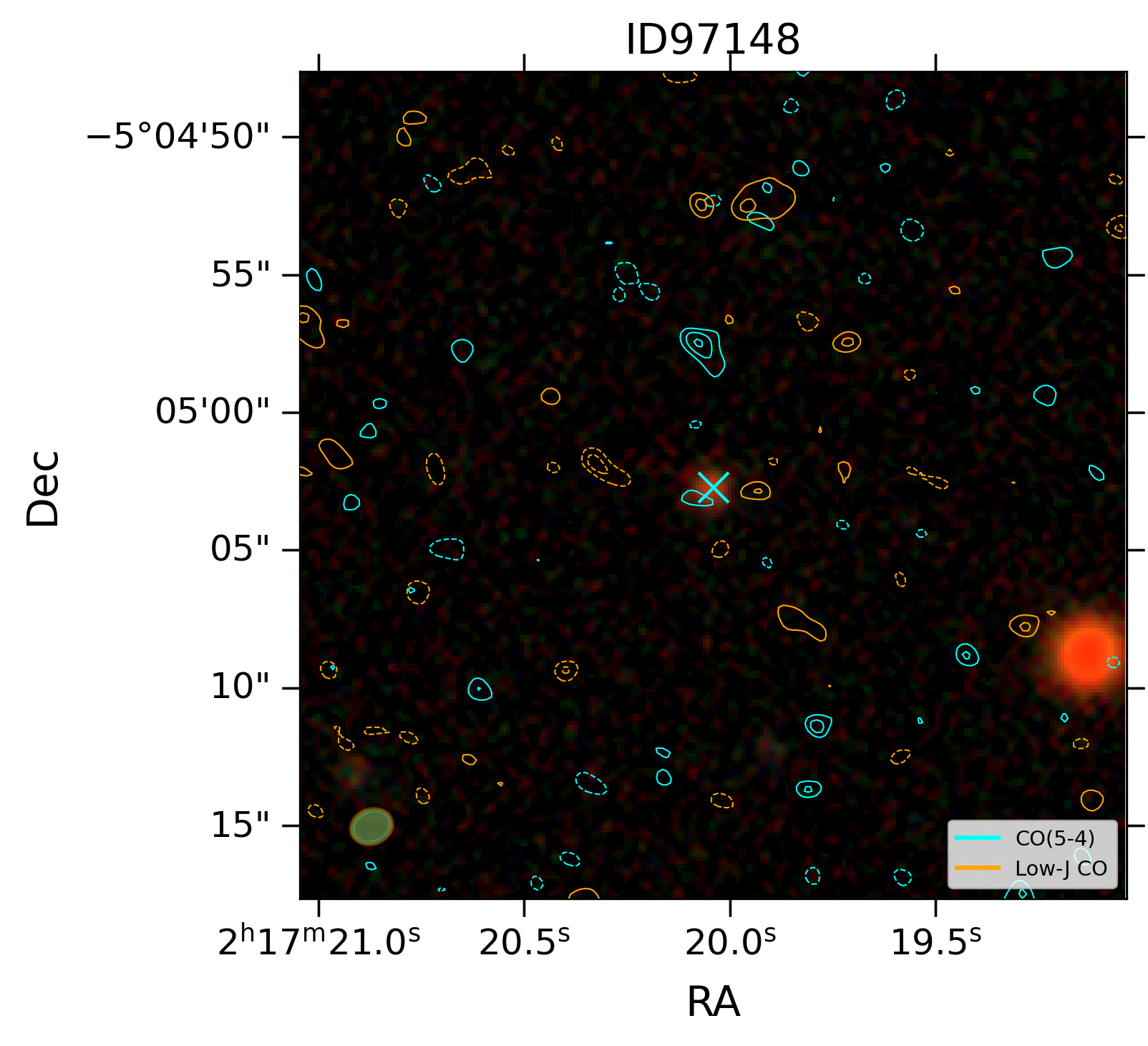}

    \caption{ALMA data of our galaxies (continued). As reported in \cite{Bezanson2021}, the $\sim 1\arcsec$ offset of the CO(2-1) of J2202 from the optical centroid is not significant given the data resolution and S/N. However, we note that the optical image of this galaxy appears to be slightly asymmetric.}
\end{figure*}

\begin{table*}
    \centering
    \caption{Measurements for our sample galaxies.} 
    \small
    \begin{tabular}{c c c c c c c c c}
    \toprule
    \midrule
    ID & RA & DEC & $z_\mathrm{CO}$ & $\mathrm{S\Delta v_{CO}}$ & $\Delta$v & $\mathrm{L'_{CO}}$ & $\mathrm{F_{cont}}$ & Line \\
       & (deg) & (deg) &  & (Jy km\, s$^{-1}$) & km s$^{-1}$ & ($\mathrm{10^8 K\, km\, s^{-1} pc^{2}}$) & ($\mathrm{mJy}$) & \\
    (1) & (2) & (3) & (4) & (5) & (6) & (7) & (8) & (9) \\
    \midrule
    J1448  & 222.191344 & 10.169614 & $0.6463 \pm 0.0001$ & $0.55 \pm 0.08$ & $390 \pm 52$ & $32.6 \pm 4.8$ & $< 11.9$ & CO(2-1)\\
           &            &           & $0.6460 \pm 0.0001$ & $1.57 \pm 0.27$ & $307 \pm 46$ & $23.0 \pm 4.0$ & $< 58.1$ & CO(4-3)\\
           &            &           & $0.6465 \pm 0.0002$ & $2.62 \pm 0.54$ & $591 \pm 93$ & $24.7 \pm 5.1$ & $< 257.2$ & CO(5-4)\\
    J2258  & 344.523653 & 23.221175 & $0.7056 \pm 0.0001$ & $1.40 \pm 0.18$ & $220 \pm 24$ & $98.1 \pm 1.2$ & $12.3$ & CO(2-1)\\
           &            &           & $0.7043 \pm 0.0005$ & $2.38 \pm 0.75$ & $836 \pm 207$ & $26.7 \pm 8.4$ & $< 303.1$ & CO(5-4)\\
    ID83492 & 34.546956 & -5.147125 & $1.1394 \pm 0.0005$ & $0.23 \pm 0.09$ & $569 \pm 162$ & $18.8 \pm 7.1$ & $< 31.9$ & CO(3-2)\\
            &           &           & $1.1398 \pm 0.0001$ & $0.16 \pm 0.05$ & $129 \pm 34$ & $4.6 \pm 1.6$ & $< 49.7$ & CO(5-4)\\
    J1109  & 167.384124 & -0.667679 & $0.5937 \pm 0.0001$ & $0.64 \pm 0.06$ & $328 \pm 21$ & $31.8 \pm 2.8$ & $< 1493.8$ & CO(2-1)\\
           &            &           &                     & $< 0.33$        &              & $< 2.6$ & $< 156.7$ & CO(5-4)\\
    J1302  & 195.703844 & 10.717528 & $0.5922 \pm 0.0002$ & $0.53 \pm 0.08$ & $732 \pm 79$ & $26.3 \pm 3.8$ & $<21.8$ & CO(2-1)\\    
           &            &           &                     & $< 0.53$        &              & $< 4.2$ & $< 210.0$ & CO(5-4)\\
    J0912  & 138.178105 & 15.384847 & $0.7472 \pm 0.0001$ & $1.01 \pm 0.07$ & $367 \pm 19$ & $79.4 \pm 5.5$ & $<257.2$ & CO(2-1)\\
           &            &           & $0.7474 \pm 0.0001$ & $0.53 \pm 0.11$ & $296 \pm 49$ & $10.5 \pm 2.3$ & $< 40.8$ & CO(4-3)\\
           &            &           &                     & $< 0.31$        &              & $< 3.9$ & $< 165.7$ & CO(5-4)\\
    J2202  & 330.601428 & -0.559743 & $0.6582 \pm 0.0001$ & $0.24 \pm 0.08$ & $181 \pm 44$ & $14.6 \pm 4.7$ & $< 38.1$ & CO(2-1)\\
           &            &           &                     & $< 2.24$        &              & $< 21.9$ & $< 84.2$ & CO(5-4)\\
    ID97148  & 34.333142 & -5.084108 & $1.2727 \pm 0.0002$ & $0.09 \pm 0.05$ & $160 \pm 64$ & $9.2 \pm 4.9$ & $< 28.6$ & CO(3-2)\\
             &           &           &                     & $<0.03$         &              & $< 1.0$ & $< 37.1$ & CO(5-4)\\
    \midrule
    C1-83492$^*$ & 34.547551 & -5.148568 & $1.1374 \pm 0.0002$ & $0.79 \pm 0.10$ &  $590 \pm 57$ & $71.5 \pm 9.1$ & $86.8 \pm 20.1$ & CO(3-2)\\
             &           &           & $1.1377 \pm 0.0002$ & $0.43 \pm 0.08$ & $388 \pm 57$ & $12.4 \pm 2.4$ & $510.4 \pm 23.7$ & CO(5-4)\\
    C2-83492$^*$ & 34.550126 & -5.149099 & $1.1379 \pm 0.0001$ & $0.89 \pm 0.11$ & $432 \pm 41$ & $63.9 \pm 8.3$ & $116.8 \pm 20.6$ & CO(3-2)\\
             &           &           & $1.1380 \pm 0.0004$ & $0.30 \pm 0.10$ & $471 \pm 119$ & $8.6 \pm 2.9$ & $318.2 \pm 37.2$ & CO(5-4)\\
    \bottomrule
    \end{tabular}
    \label{tab:measurements}
    \tablefoot{Columns: (1) Galaxy ID; (2) Right ascension of the galaxy center; (3) Declination of the galaxy center; (4) Redshift estimated by fitting the emission line with a Gaussian in our 1D ALMA spectra. The uncertainty that we report is the formal error obtained from the fit; (5) CO integrated flux density. Upper limits are $3\sigma$; (6) Line velocity width; (7) CO brightness temperature. Upper limits are $3\sigma$; (8) Flux of the continuum underlying the line. Upper limits are $3\sigma$; (9) CO transition.
    \\ $^*$ Indicate star-forming companions of ID83492.}
\end{table*}

\begin{table*}[t]
\centering
\caption{Physical properties of our sample galaxies.}
\small
\begin{tabular}{c c c c c c}
\toprule
\midrule
ID & log(M$_\star$/M$_\odot$) & SFR & log(M$_\mathrm{H2}$/M$_\odot$) & sSFR & $R_{ij}$ \\
   &                          & ($\mathrm{M_\odot\, yr^{-1}}$) &               & ($\mathrm{Gyr^{-1}}$) & \\
(1) & (2) & (3) & (4) & (5) & (6) \\
\midrule
J1448$^a$ & $11.6^{+0.04}_{-0.07}$ & $1.06^{+0.99}_{-0.94}$ & $10.4\pm 0.06$ & 0.003 & $	0.76 \pm 0.19$ \\
J2258$^a$ & $11.8^{+0.03}_{-0.05}$ & $0.94^{+1.85}_{-1.71}$ & $10.9\pm 0.05$ & 0.001 & $0.27 \pm 0.09$\\
ID83492$^b$ & $10.8^{+0.20}_{-0.20}$ & $0.52^{+0.79}_{-0.32}$ & $10.4\pm 0.16$ & 0.008 & $0.24 \pm 0.12^c$ \\
J1109$^a$ & $11.3^{+0.09}_{-0.03}$ & $2.33^{+1.12}_{-1.62}$ & $10.4\pm 0.04$ & 0.012 & $< 0.08$ \\
J1302$^a$ & $11.6^{+0.04}_{-0.06}$ & $0.26^{+0.93}_{-0.26}$ & $10.4\pm 0.06$ & 0.001 & $< 0.16$ \\
J0912$^a$ & $11.2^{+0.03}_{-0.02}$ & $0.81^{+1.33}_{-0.76}$ & $10.8\pm 0.03$ & 0.003 & $< 0.05$ \\
J2202$^a$ & $11.2^{+0.03}_{-0.02}$ & $1.99^{+1.91}_{-1.70}$ & $10.1\pm 0.14$ & 0.004 & $< 1.50$ \\
ID97148$^b$ & $10.7^{+0.10}_{-0.10}$ & $< 0.72$               & $10.1\pm 0.23$ & $< 0.016$ & $< 0.11^c$ \\
\midrule
C1-83492$^b$ & $10.6 \pm 0.1$ & $295.12^{+217.74}_{-225.94}$ & $9.9\pm 0.06$ & 7.9 & $0.19 \pm 0.04^c$ \\
C2-83492$^b$ & $10.8 \pm 0.20$ & $851.14^{+19.83}_{-651.61}$ & $9.9\pm 0.06$ & 13.5 & $0.12 \pm 0.04^c$ \\
\bottomrule
\end{tabular}
\label{tab:properties}
    \tablefoot{Columns: (1) ID; (2) Stellar mass from optical SED fit. We report the formal uncertainties from the SED modelling, although more realistic uncertainties accounting for systematics are $\sim 0.2\, \mathrm{dex}$ \citep{Pacifici2023}; (3) Star formation rate from optical SED; (4) Molecular gas mass. We used $\alpha_\mathrm{CO} = 4.4\, \mathrm{M_\odot pc^{-2} (K\, km\, s^{-1})^{-1}}$ for the post-starbursts and $\alpha_\mathrm{CO} = 0.8\, \mathrm{M_\odot pc^{-2} (K\, km\, s^{-1})^{-1}}$ for the starbursting companions; (5) Specific star formation rate; (6) Ratio of the CO(5-4) and lower-$J$ brightness temperature. In this table we report R$_{53} = L'_\mathrm{CO(5-4)}/L'_\mathrm{CO(3-2)}$ for ID83492 and ID97148, while we report R$_{52} = L'_\mathrm{CO(5-4)}/L'_\mathrm{CO(2-1)}$ for the rest of the sample.
    \\ $^a$Stellar mass and SFR from \cite{Bezanson2021}. $^b$Stellar mass and SFR from \cite{Zanella2023}. $^c$We report the R$_{53}$ ratio, while in the plots we adopt a pseudo-R$_{52}$ ratio estimated adopting R$_{32} = 0.62$ (average of local post-starbursts value, \citealt{French2023}).}
\end{table*}

\section{Data}
\label{sec:data}
We focus on 8 gas-rich post-SB galaxies at $0.6 < z < 1.3$. They were all the publicly available (to our best knowledge) and observable by ALMA $z \gtrsim 0.6$ post-SBs with detected low-$J$ CO ($J=2,3$) lines at the time when the observations were proposed \citep{Suess2017, Bezanson2021, Zanella2023}.

\subsection{Sample selection and ancillary data}
\label{subsec:target}
The sample is extracted from two surveys: SQuIGG$\vec{L}$E \citep{Suess2022} and the study of \cite{Zanella2023}. 

The SQuIGG$\vec{L}$E sample is selected from the Sloan Digital Sky Survey DR14 spectroscopic database \citep{Abolfathi2018} to have strong Balmer breaks, blue slopes redward of the break, and $z \sim 0.7$. For a description of the spectroscopic identification, stellar populations, and physical characterization of SQuIGG$\vec{L}$E galaxies we refer the reader
to the survey paper \citep{Suess2022}. A subset of the SQuIGG$\vec{L}$E sample has been observed with ALMA targeting the CO(2-1) line emission \citep{Suess2017, Bezanson2021}. Of the 13 targeted galaxies, CO(2–1) emission was detected in six. These 6 galaxies are those we focus on in this study. We note that a more recent study targeted the CO(2-1) emission of 50 additional post-SBs detecting 27 of them \citep{Setton2025}. None of them have so far published data of higher-$J$ CO transitions. However, our sample selection and observations were concluded before this study was published.

The galaxies from \cite{Zanella2023} were initially optically selected by \cite{Wild2016} applying a principal component
analysis on their spectral energy distributions (SED) based on the photometry of the Ultra-Deep Survey \citep{Lawrence2007}. The post-SB nature of a subsample of 19 objects with deep absorption lines
(EW(H$\delta$)$> 5$\AA) and minimal  [OII]3727$\mathrm{\AA}$ emission consistent with the unobscured SFR estimated from the UV luminosity was confirmed by \cite{Maltby2016}. Two $z > 1$ targets from this subsample were targeted with ALMA by \cite{Zanella2023} and the CO(3-2) emission line was detected.

In Table \ref{tab:log} we report the ALMA observations for our sample.

\subsection{ALMA CO(5-4) data}
\label{subsec:ALMA_data}

We carried out ALMA Band 6 (for $z < 1$ sources) or 7 (for $z > 1$ sources) observations for our sample during Cycle 11 (Program 2024.1.00061, PI: A. Zanella) with the goal of detecting the CO(5-4) emission line at rest-frame frequency $\nu_\mathrm{rf} = 576.267$ GHz and the underlying continuum, redshifted in the frequency range $\nu_\mathrm{obs} = 260 - 360$ GHz.

Each source was observed for 8-110 minutes. The sensitivities achieved are reported in Tab. \ref{tab:log}. The native spectral resolution of the observations is 6.5 - 13.5 km s$^{-1}$, later binned to three times lower velocity resolution for our purposes. The angular resolution of the observations is in the range $0.5\arcsec - 1.4\arcsec$ (Tab. \ref{tab:log}).

The data were reduced with the standard ALMA pipeline, based on the CASA software \citep{McMullin2007}. The calibrated data cubes were then converted to \textit{uvfits} format and analyzed with the software GILDAS \citep{Guilloteau2000}. For consistency, we also retrieved from the ALMA archive the publicly available low-$J$ CO data and reduced them adopting the same procedure of our CO(5-4) observations.

\section{Analysis}
\label{sec:analysis}

\subsection{CO(5-4) emission}
\label{subsec:co54}
To produce the velocity-integrated CO(5-4) line maps for our targets, we first determined the optimal spectral range over which to integrate the spectra. This was done through an iterative procedure, following an approach similar to that described in \cite{Zanella2023}. We modeled the emission in the \textit{uv} plane\footnote{We performed the entire analysis, including flux measurements, in the \textit{uv} plane. Being based on the raw interferometric visibilities, it does not require image reconstruction or deconvolution, avoiding biases and artefacts introduced during imaging and cleaning, and allowing us to fully account for the spatial frequencies sampled by the array, as well as the instrumental resolution and sensitivity.} as an elliptical Gaussian, fitting channel by channel across all four sidebands with the GILDAS task \texttt{uv\_fit}. Our sources are expected to be unresolved and the size of the elliptical Gaussian is indeed consistent with the beam size. Nevertheless, we chose to adopt an extended model rather than a point source, to ensure consistency with the analysis of the low-$J$ CO emission, which in several cases is spatially resolved (Section \ref{subsec:co21}). We verified that using a point-source model yields CO(5–4) integrated fluxes and upper limits that are, on average, 1.5 times fainter. However, this difference does not affect our conclusions.

The spatial positions of the models were fixed to those derived from the CO(2–1) emission, which are consistent with the positions measured from optical imaging. Using these models, we extracted one-dimensional spectra and searched for positive line emission. When such a signal was detected, we averaged the channels that maximized the signal-to-noise ratio (S/N), and used the resulting channel-averaged two-dimensional map to fit the spatial position of the line. We considered a galaxy to be detected if its peak flux reached a significance of at least $3\sigma$. In addition, we verified that the redshift of the detected CO line is consistent with the spectroscopic redshift from optical data and that the CO emission is spatially consistent with the optical and IR emission. We estimated the total line flux by fitting a 1D Gaussian to the extracted optimal S/N spectrum.
When a line was not detected, we instead created two-dimensional maps using the channel range identified for the CO(2–1) emission and adopted a $3\sigma$ flux upper limit.
We securely detected the CO(5-4) emission line for 3 galaxies, while the remaining are undetected (Fig. \ref{fig:sample}). 

To increase the signal-to-noise ratio for non detections we aligned the individual CO(5-4) maps to the CO(2-1) positions and stacked them (coadding visibilities). Even the stacking yielded a non detections. We estimated a CO(5-4) flux upper limit by fitting the stacked data in the $uv$ plane with the Fourier transform of a 2D Gaussian with parameters fixed to those obtained by stacking the CO(2-1) stacking.  

We estimated the redshift of the 3 detected CO(5-4) lines in two ways, both giving consistent results ($\delta z< 0.001$): by computing the signal-weighted average frequency within the line channels and by fitting the one-dimensional spectrum with a Gaussian function. We compared these redshift estimates with those obtained from CO(2-1) \citep{Bezanson2021, Zanella2023} and found that they agree within the uncertainties.
The CO(5-4) total fluxes and redshifts are reported in Table \ref{tab:measurements}.

\subsection{Low-$J$ CO emission}
\label{subsec:co21}
Since the main goal of this study is to compare the CO(5–4) and lower-$J$ CO transitions, we re-analysed archival public data targeting the CO(2–1), CO(3–2), and/or CO(4–3) emission lines using the same procedure adopted for the CO(5–4) emission (Section \ref{subsec:co54}). The fluxes we obtain are on average 20\% fainter than those reported in the literature \citep{Suess2017, Bezanson2021, Zanella2023}, due to the different methodology used to extract the one-dimensional spectra and measure fluxes. In our analysis, we adopt fixed elliptical apertures that are held constant across all velocity channels and across all datasets. In contrast, \citet{Suess2017} and \citet{Bezanson2021} adopt circular Gaussian models with radii that vary on a channel-by-channel basis. We prefer the fixed-aperture approach because it minimizes the impact of noise fluctuations in individual channels, and it ensures that exactly the same spatial region is used when measuring fluxes in different transitions, which is particularly important for robust line ratio measurements. We verified that using elliptical instead of circular Gaussian models yields negligible differences. The main difference with respect to the \cite{Zanella2023} dataset instead lies in the adopted source model. While \cite{Zanella2023} use a point-source model, motivated by the fact that the CO(3–2) transition appeared unresolved in their data, we adopt an elliptical Gaussian model. This choice is made to ensure methodological consistency with the analysis applied to the rest of our sample. Since the same apertures are applied consistently to all datasets, any systematic difference in the absolute flux normalization does not affect the derived line ratios.
We report all fluxes in Table \ref{tab:measurements}. A comparison of the spatial locations and morphologies of the different emission lines is shown in Figure \ref{fig:sample}.

\subsection{Continuum emission}
\label{subsec:continuum}
We generated averaged continuum maps by integrating over the spectral range. For the CO(5–4) detected galaxies, we excluded the channels dominated by line emission, while for the non-detections we excluded the channels where emission is expected based on the low-$J$ CO transitions (see Section \ref{subsec:co21} for details). None of our main targets is detected in the $\nu_\mathrm{obs} = 260$–$360$ GHz continuum, down to sensitivities in the range $\mathrm{12.6 - 99.3\, \mu Jy}$.
We also produced continuum maps by averaging over a broader spectral range, including not only the continuum underlying the CO(5–4) line but also that associated with the lower-$J$ transitions. This analysis likewise yielded no detections, down to sensitivities in the range $\mathrm{4.5 - 21.0\, \mu Jy}$. Since no detections were found, we did not refine this analysis further, for example by assuming a continuum slope when stacking across different frequency ranges.

\subsection{Detection of companion galaxies}
\label{subsec:neighbours}

Three of our sample galaxies have companions or tidal features that have been identified in low-$J$ CO maps. We verified whether they are also detected in CO(5-4).

In particular, \cite{Zanella2023} serendipitously detected bright CO(3-2) emission lines ($14.0\sigma$ and $16.5 \sigma$) and continuum ($3.5\sigma$ and $4.7\sigma$) from two galaxies in the surroundings of ID83492. Those galaxies are also detected in CO(5-4), despite having lower significance ($10.6 \sigma$ and $5.5 \sigma$). They both have optical counterparts in \textit{Hubble Space Telescope} Wide Field Camera 3 (\textit{HST}/WFC3) imaging from the 3D-HST program (\citealt{Skelton2014}) and their CO(5-4) redshifts are $z = 1.1377 \pm 0.0002$ and $z = 1.1380 \pm 0.0004$, consistent with the CO(2-1) redshift ($\mathrm{\Delta z \lesssim 0.0003 }$). Given their velocity offset ($v_\mathrm{off} \sim 250\, \mathrm{km\, s^{-1}}$ and $v_\mathrm{off} \sim 195\, \mathrm{km\, s^{-1}}$) and projected distance from ID83492 ($5.8\arcsec \sim 48\,\mathrm{kpc}$ and $13.7\arcsec \sim 112\,\mathrm{kpc}$), they are likely companions that might merge with our target galaxy within $< 2.5$ Gyr, as already reported by \cite{Zanella2023}. 
We report the properties of all companions in Table \ref{tab:measurements} and \ref{tab:properties}. 

Additionally, \cite{Spilker2022} and \cite{Donofrio2025} have analyzed spatially resolved CO(2-1) maps of the J1448 and J2258 galaxies, revealing extended molecular gas tidal tails spanning up to 65 kpc, a clear indication of recent mergers. Such extended tidal features are not detected in CO(5-4) likely due to the lack of sufficient sensitivity. Only a tentative ($\sim 4\sigma$) CO(5-4), spatially unresolved, emission emerges at the location of the Northern tail N3 identified by \cite{Donofrio2025} with CO(2-1) and H$\alpha$ emission. Its CO(5-4) redshift ($z = 0.6457 \pm 0.0003$) is consistent with that of the central body of J1448 (Table \ref{tab:measurements}).

Finally, we searched for dust continuum detections nearby our targets. We report the detections in Appendix \ref{app:continuum_det}, while coordinates and fluxes are reported in Table \ref{tab:continuum}.

\subsection{Physical properties}
\label{subsec:phys_prop}

Stellar masses and SFR were obtained by modeling multi-wavelength photometry as described in \cite{Bezanson2021} and \cite{Zanella2023}. In brief, \cite{Bezanson2021} modelled the photometry and spectra simultaneously using \texttt{Prospector} \citep{Leja2017} with a custom set of
``non parametric'' star formation histories, assuming a
\cite{Kriek2013} dust law\footnote{They also
fit the spectra and photometry with delayed exponential
SFHs \cite{Setton2020} assuming similar
\cite{Chabrier2003} IMF, \cite{Bruzual2003} libraries, and a
\cite{Calzetti2000} dust law using \textsc{FAST++}. The stellar masses derived from these fits are an average of 0.38 dex lower than the stellar masses
derived in the default fits.}. \cite{Zanella2023} modelled the photometry and the spectra simultaneously with \textsc{FAST++}\footnote{\href{https://github.com/cschreib/fastpp}{https://github.com/cschreib/fastpp}} \citep{Schreiber2018} using \cite{Bruzual2003} models, the \cite{Chabrier2003} IMF, delayed SFHs, the \cite{Calzetti2000} dust
attenuation curve, and fixed the
redshifts to the spectroscopic values. We report the specific star formation rate $\mathrm{sSFR = SFR/M_\star}$ in Table \ref{tab:properties}.

The molecular gas masses ($\mathrm{M_{H2}}$) were estimated from the CO(2-1) emission for the galaxies at $z < 1$ \citep{Bezanson2021} and from the CO(3-2) line for the targets with $z > 1$ \citep{Zanella2023}. Both \citep{Bezanson2021} and \citep{Zanella2023} use a Milky Way-like CO-to-H2 conversion factor \citep{Bolatto2013} and adopt the following excitation factors to convert the higher $J$ transitions to CO(1-0): $R_{31} = 0.5$ \citep{Carilli2013} and $R_{21} = 1$ \citep{Combes2007, Dannerbauer2009, Young2011}. They obtain molecular gas fractions in the range $1\% - 30\%$. We re-estimated the molecular gas masses considering the average excitation factors obtained for local post-SBs, namely $R_{31} = 0.32$, $R_{21} = 0.52$, and $R_{32} = 0.62$ \citep{French2021}. We obtain on average 0.2~dex larger gas masses and gas fractions in the range $2\% - 40\%$ (Table \ref{tab:properties}). However, since gas masses and fractions play only a minor role in this study, our conclusions are not affected by this choice.

For the two starbursting companions of ID83492 instead we considered a CO-to-H2 conversion factor and excitation ratios of typical submillimeter galaxies, namely $R_{31} = 0.66$, $R_{21} = 0.85$, $R_{32} = 0.78$ and $\alpha_\mathrm{CO} = 0.8\, \mathrm{M_\odot pc^{-2} (K\, km\, s^{-1})^{-1}}$ \citep{Carilli2013, Bolatto2013}.

All our targets have available estimates of the D$_n$4000 spectral index, a useful tool to constrain the mean stellar age of galaxies with younger galaxies having on average a less pronounced 4000-break \citep[reported in][]{Bezanson2021, Zanella2023}. For the SQuIGG$\vec{L}$E sample \cite{Bezanson2021} and \cite{Suess2021} also published an estimate of the time since quenching derived from the star formation history provided by the \texttt{Prospector} stellar population synthesis modeling.

\section{Results}
\label{sec:results}
Observations of multiple CO transitions provide a diagnostic of the physical conditions of the interstellar medium. The CO line ratios and the shape of the SLED are sensitive to the gas volume density, kinetic temperature, and column density, thereby constraining the properties of the molecular gas in galaxies \citep{Young1991, Solomon2005, Carilli2013, Narayanan2014, Bournaud2015, Popping2016, Kamenetzky2018, Combes2018, Renaud2019a, Renaud2019b, Valentino2020b, Liu2021}. 

\begin{figure}
    \centering
    \includegraphics[width=0.5\textwidth]{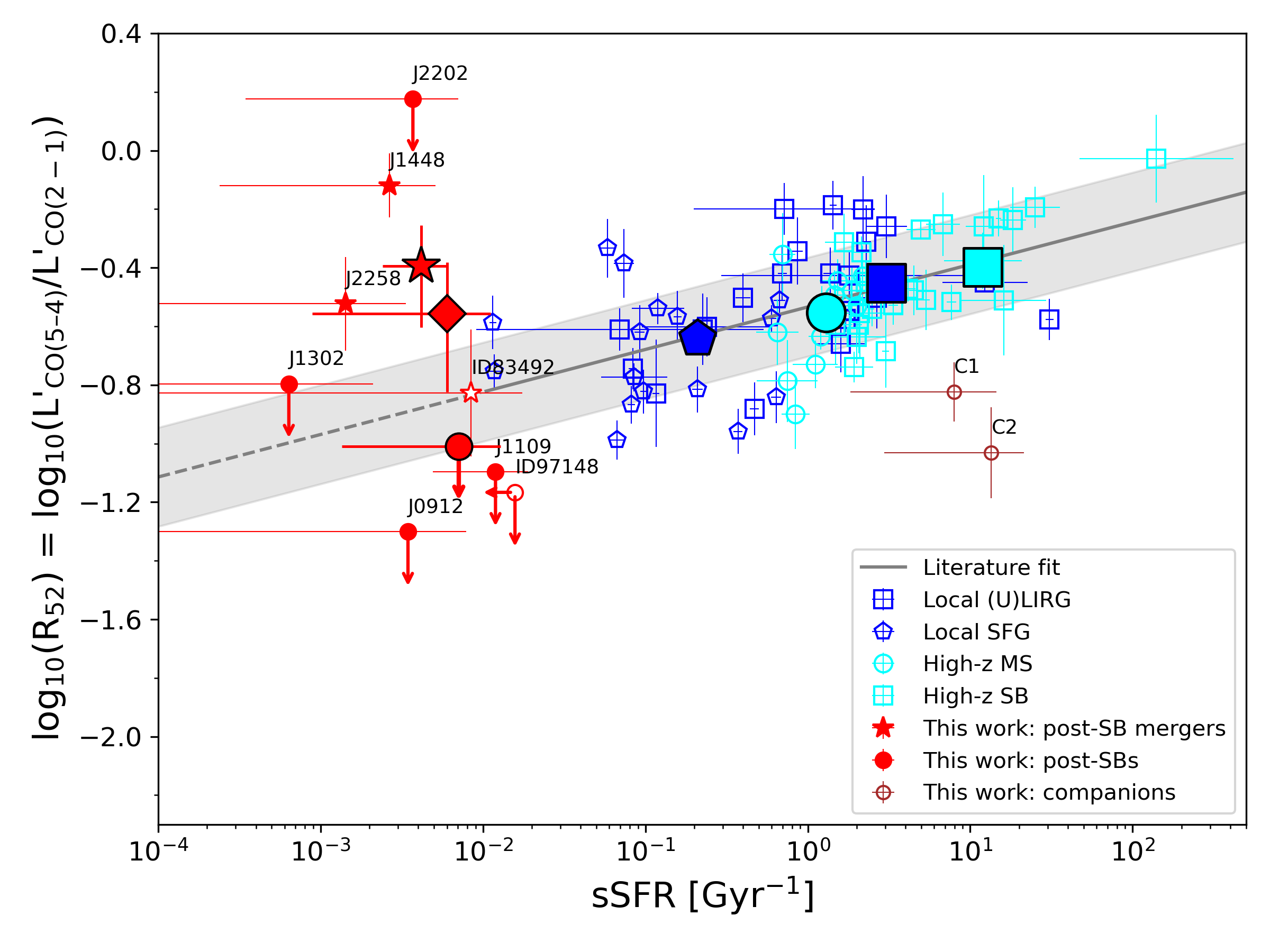}
    \caption{Ratio of CO(5-4) to CO(2-1), a proxy for gas excitation, as a function of specific star formation rate. Our sample is shown with red circles (non mergers) and stars (mergers) and is compared to literature samples of local star-forming galaxies (blue pentagons), (U)LIRGs (blue squares), high-redshift main-sequence galaxies (cyan circles), and high-redshift SBs (cyan squares; \citealt{Liu2021}). The companions of ID83492 are shown as brown circles. For post-SBs without CO(2-1) observations, empty symbols indicate a pseudo CO(2-1) estimated from the CO(3-2) assuming the average local post-SBs $R_{32} = 0.62$ ratio \citep{French2023}. Large symbols indicate the population averages (see Sec. \ref{subsec:excitation}). A fit to the literature samples (gray solid line) with its scatter (shaded area) from \cite{Valentino2020b} is shown and extrapolated toward lower sSFR (gray dashed line).}
    \label{fig:co_ratio_ssfr}
\end{figure}

\begin{table}[t!]
    \centering
    \caption{Mean specific star formation and $R_{52}$ ratio for our sample.}
    \small
    \begin{tabular}{c c c}
    \toprule
    \midrule
                      & sSFR & R$_{52}$ \\
                      & (Gyr$^{-1}$) & \\
                      \midrule
    All post-SBs & 0.006 & $0.28^{+0.15}_{-0.13}$ \\
    Merging post-SBs & 0.004 & $0.40^{+0.15}_{-0.15}$ \\
    Non merging post-SBs & 0.007 & $< 0.10$ \\
    Companions ID83492 & 10.716 & $0.12$ \\
    \bottomrule
    \end{tabular}
    \tablefoot{The R$_{52}$ is estimated as an average in case of CO(5-4) detections, namely the merging post-SBs and the companions of ID83492. The uncertainty for the merging post-SBs is estimated through bootstrap, while we do not estimate an uncertainty for the companions as they are only two. For the CO(5-4) non detections $R_{52}$ is estimated through stacking (Sec. \ref{subsec:excitation}). For the entire population we estimated the mean ratio using the Kaplan–Meier estimator from survival analysis to account for measurements and upper limits (Sec.~\ref{subsec:excitation}, with uncertainties estimated through bootstrap. For ID83492 and ID97148 we estimated the CO(2-1) flux assuming $R_{32} = 0.62$, while for their starbursting companions we adopted $R_{32} = 0.78$ (Sec. \ref{subsec:excitation}).}
    \label{tab:ratios}
\end{table}

\subsection{Molecular gas excitation}
\label{subsec:excitation}

By combining CO(5–4) and lower-$J$ CO transitions, we investigated the ratios $R_{52} = L'\mathrm{CO(5-4)}/L'\mathrm{CO(2-1)}$ and $R_{53} = L'_\mathrm{CO(5-4)}/L'_\mathrm{CO(3-2)}$, which serve as proxies for the excitation of the molecular gas. In Figure \ref{fig:co_ratio_ssfr}, we show $R_{52}$ as a function of sSFR for our sample. For two galaxies, ID83492 and ID97148, CO(2–1) observations are not available. In these cases, we estimated $L'_\mathrm{CO(2-1)}$ from $L'_\mathrm{CO(3-2)}$ assuming the average ratio $R_{32} = 0.62$ of local post-starbursts \citep{French2023}.

We also include the two star-forming companions of ID83492, as well as comparison samples of local and high-redshift main-sequence galaxies, local Ultra Luminous Infrared Galaxies (ULIRGs), and high-redshift starbursts from the literature \citep{Liu2021}. The post-SBs extend the region of parameter space sampled by star-forming galaxies toward lower sSFR. 
All our sample galaxies are detected in CO(2-1), while only 3 are detected in CO(5-4). Non-detections have on average a tight upper limit $R_{52} < 0.10$ ($1\sigma$) derived by stacking (see Sec. \ref{subsec:co54}). This is twice smaller than the average for local star-forming galaxies, and 2.6–3.7 times lower than the $R_{52}$ of (U)LIRGs, high-redshift SBs and main-sequence galaxies. This suggests a reduced fraction of dense, warm gas relative to cold, diffuse molecular gas.

The three CO(5–4) detections in our sample instead have, on average, a higher $R_{52} = 0.40$, more comparable to that of high-redshift starbursts, than the rest of the sample. We note that only the  CO(5–4) detections in our sample correspond to mergers, while galaxies without detections show no evidence of companions or merger signatures (e.g., tidal tails). Their relatively high $R_{52}$ ratio suggests that quenching is still ongoing in these systems, with star formation likely depleting the remaining molecular gas on short timescales. Such behavior indicates that gas exhaustion and potentially gas stripping are important mechanisms to keep galaxies quiescent.
Interestingly, the two star-forming companions of ID83492 have sSFRs comparable to those of starburst galaxies, but their $R_{52}$ values are lower by about 0.5 dex and comparable to the most stringent upper limits in our sample. This suggests that the ongoing merger has affected the molecular gas properties of all the galaxies involved, and that a mismatch between the timescales traced by $R_{52}$ and sSFR is key to interpreting these results (see Section \ref{sec:discussion}).

Finally, we computed the average $R_{52}$ ratio of the entire sample by employing the Kaplan–Meier estimator from survival analysis to properly account for both detections and upper limits\footnote{We used the python package \texttt{lifelines} and the task \texttt{KaplanMeierFitter}}. We find and average $R_{52} = 0.28$ (Table~\ref{tab:ratios}). To estimate the associated uncertainties we employed a bootstrap resampling approach.

We also investigated if the $R_{52}$ ratio correlates with other physical properties, such as the depth of the 4000 Å break ($D_n4000$\AA), the time since quenching ($t_\mathrm{q}$), or the molecular gas fraction ($f_\mathrm{H2}$). We do not find any clear correlation with $D_n4000$\AA or $t_\mathrm{q}$, and a mild trend with $f_\mathrm{H2}$ (Fig. \ref{fig:co_ratio_properties}). However, a larger sample with more detections is needed to confirm it.

In Table \ref{tab:ratios} we report the $R_{52}$ of our sample.

\begin{figure}[t!]
    \centering
    \includegraphics[width=\linewidth]{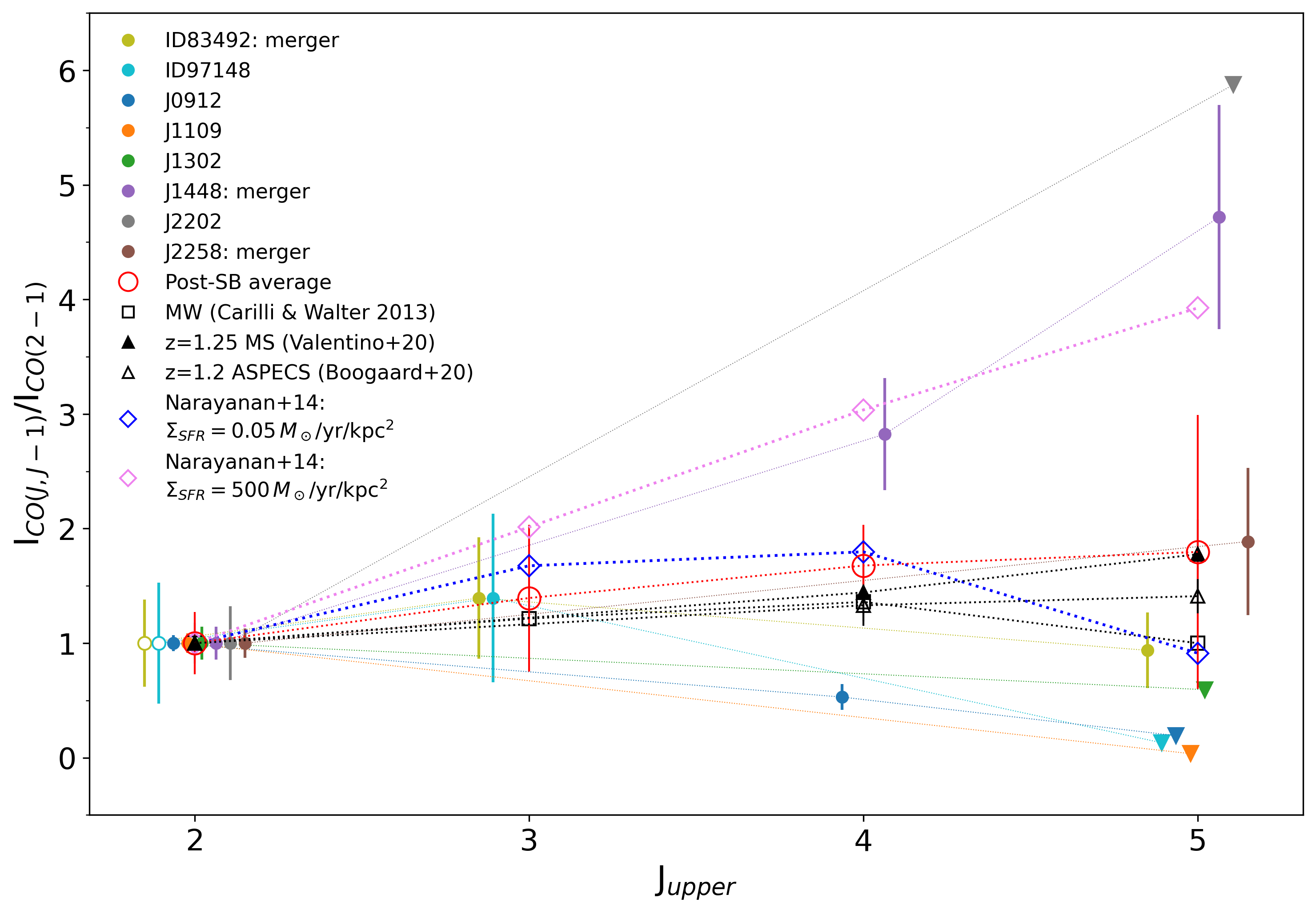}
    \caption{CO SLED of our post-starburst galaxies flux ratio. Individual galaxies are shown with small colored circles (measurements) and down-pointing triangles (3$\sigma$ upper limits), while the average SLED is indicated by red empty circles. For comparison, we also show the Milky Way SLED (black squares; \citealt{Carilli2013}), the average SLED of main-sequence galaxies (black filled triangles from \citealt{Valentino2020b} and black empty triangles from \citealt{Boogaard2020}), and the SLED predicted by the model of \cite{Narayanan2014} assuming a SFR surface density of $\Sigma_\mathrm{SFR} = 0.05\, \mathrm{M_\odot\, yr^{-1}\, kpc^{-2}}$ (blue diamonds) and $\Sigma_\mathrm{SFR} = 500\, \mathrm{M_\odot\, yr^{-1}\, kpc^{-2}}$ (pink diamonds).}
    \label{fig:sleds}
\end{figure}

\subsection{Spectral line energy distribution}
\label{subsec:sled}

The CO SLED traces how the radiated energy is distributed among the different $J$ transitions of the CO ladder and provides a powerful diagnostic of the physical conditions of the ISM. It has been shown that the CO SLED can distinguish between galaxies with low SFRs, such as the Milky Way, whose emission peaks at intermediate $J = 3{-}4$ \citep{Weiss2005, Carilli2013}, and dense, warm starbursts, whose emission peaks at higher $J = 7{-}8$ \citep{Weiss2007, Vallini2019, Esposito2024}. In the presence of active AGN, even higher excitation is observed due to the contribution of X-ray dominated regions \citep{vanderWerf2010}. Both observational \citep{Papadopoulos2012, Daddi2015, Valentino2020b, Liu2021} and theoretical studies \citep{Narayanan2014, Bournaud2015, Vollmer2017} have shown that high-redshift star-forming galaxies have higher molecular gas fractions, denser gas, and enhanced SFRs \citep{Daddi2010, Genzel2010, Tacconi2018}, resulting in SLEDs that peak at higher $J = 4{-}5$ compared to local star-forming galaxies.

By complementing our observations with archival CO(2-1), CO(3-2), and CO(4-3) data when available, we investigated the CO SLEDs of post-starbursts and compared them with literature observational and theoretical results. In Figure \ref{fig:sleds}, we show the individual SLEDs of our sample galaxies, all normalized to the CO(2-1) transition. As before, for the two galaxies lacking CO(2-1) observations, we derived a pseudo CO(2-1) using $R_{32} = 0.50$ (see Section \ref{subsec:excitation}). We find a diversity of SLED shapes among our post-starbursts, with some peaking at $J \geq 5$ (e.g., J1448, J2258) and others peaking at $J = 3$ (e.g., ID97148, J0912). The average SLED of our targets peaks at $J = 4$ and exhibits a shape comparable to the Milky Way SLED \citep{Carilli2013}.

The emission of two of the three confirmed mergers in our sample peaks at $J \geq 5$, similar to what is observed in high-redshift main-sequence galaxies \citep{Valentino2020b, Boogaard2020}. For the remaining merger, only two data points are available (CO(3-2) and CO(5-4)), so additional observations are required to fully constrain the shape of its SLED.

Finally, we compared our observations with the model proposed by \cite{Narayanan2014}, which links a galaxy’s mean SFR surface density ($\Sigma_\mathrm{SFR}$), a proxy for gas density and temperature, to the shape of its SLED. We show in Fig.~\ref{fig:sleds} the SLED predicted for $\Sigma_\mathrm{SFR} = 0.05\, \mathrm{M_\odot\, yr^{-1}\, kpc^{-2}}$, consistent with the average $\Sigma_\mathrm{SFR}$ of our sample galaxies computed using their optical size. This model reproduces the shape of the average SLED of our post-SBs, peaking at $J \sim 4$ and declining at higher $J$.

To reproduce the SLEDs of J1448 and J2258, higher $\Sigma_\mathrm{SFR}$ values are required. In particular, models with $\Sigma_\mathrm{SFR} \sim 300 - 500\, \mathrm{M_\odot\, yr^{-1}\, kpc^{-2}}$ are needed to reproduce the rising SLED of J2258 up to $J = 5$, similar to high-redshift starbursts, although additional data points at intermediate $J$ (e.g., CO(3-2), CO(4-3)) are needed to fully constrain its shape. Instead even models with very high SFR surface densities ($\Sigma_\mathrm{SFR} > 500\, \mathrm{M_\odot\, yr^{-1}\, kpc^{-2}}$) struggle to reproduce the SLED of J1448, which rises steeply toward higher $J$ transitions. The extreme shape of this SLED could be due to the presence of an AGN and/or shocks \citep{Narayanan2014}. Indeed, the optical spectrum of J1448 shows luminous and broad [OIII] and H$\alpha$ emission lines, indicative of AGN activity \citep{Greene2020, Donofrio2025, Zhu2025}, and VLA 6 GHz observations reveal a morphology consistent with compact, young radio jets \citep{Donofrio2025}. Additionally, shocks from AGN outflows and/or the ongoing merger may further enhance the CO excitation.

For the other two mergers in our sample, evidence for AGN activity is less clear. In J2258, \cite{Donofrio2025} report only faint H$\alpha$ emission and compact, weak radio emission consistent with past star formation. Optical and radio data for ID83492 are not available; however, it is covered by deep \textit{Chandra} observations\footnote{Among our sources, only ID83492 and ID97148 are covered by deep \textit{Chandra} observations, which did not detect them \citep{Evans2024}.}, which also yielded no detection \citep{Evans2024}.

\subsection{Residual star formation}
\label{subsec:residual_sf}

Some works have found that a non negligible fraction ($\sim 25\%$) of local gas-rich post-SBs host obscured star formation at levels consistent with those observed in star-forming galaxies and starbursts, thereby questioning their truly quiescent nature \citep{Poggianti2000, Smercina2018, Baron2022, Baron2023}.
Interestingly, our galaxy J1448, which exhibits a 60 kpc-long CO(2-1) tidal tail \citep{Spilker2022} due to an ongoing merger, also shows pockets of star formation traced by H$\alpha$ in the northern tail, as revealed by \textit{HST}/WFC3 spectroscopy \citep{Donofrio2025}. One of these H$\alpha$-emitting regions also emits CO(5-4), indicating a dense molecular gas clump actively forming stars. H$\alpha$ and [OIII] are also detected in the galaxy core; however, these are consistent with AGN activity \citep{Zhu2025, Donofrio2025}, suggesting that the core is transitioning toward quiescence. Faint H$\alpha$ is reported in the core of J2202 as well, though it is unclear if it originates from residual star formation or AGN \citep{Donofrio2025}. Finally, ID83492 and ID97148 show [OII], which may arise from residual star formation (dust-corrected $\mathrm{SFR = 0.2\, M_\odot\, yr^{-1}}$ for ID97148 and $\mathrm{0.3\, M_\odot\, yr^{-1}}$ for ID83492; \citealt{Zanella2023}) or AGN activity.

We used the CO(5-4) - $L_{\rm IR}$ relation calibrated on star-forming galaxies as a consistency test to assess whether the CO(5-4) line luminosities observed in our sample could plausibly arise from obscured star formation.
We converted the CO(5-4) luminosity into total IR luminosity ($\mathrm{L_{IR}}$) following the relation of \cite{Valentino2020b}, and then converted $\mathrm{L_{IR}}$ into SFR using the relation from \cite{Kennicutt2021}. While undetected galaxies have relatively low SFRs ($\mathrm{SFR \lesssim 20\, M_\odot\, yr^{-1}}$), the three sources with detected CO(5-4) emission would imply significantly higher obscured SFRs ($\mathrm{SFR \sim 20 - 140\, M_\odot\, yr^{-1}}$). We verified whether such high IR luminosities are consistent with the non-detections in the dust continuum. To this end, we used the average SED template for quiescent galaxies from \cite{Magdis2021}, which is characterized by a dust temperature $T_\mathrm{d} = 20\, \mathrm{K}$, and we rescaled it to match the most stringent dust continuum $3\sigma$ upper limit available for each galaxy. This yielded the maximum $\mathrm{L_{IR}}$ allowed by the lack of continuum detection, and maximum obscured $\mathrm{SFRs < 18\, M_\odot\, yr^{-1}}$, consistent with values expected for quiescent galaxies. 
However, \cite{Valentino2020b} calibrated the CO(5-4) - L$_\mathrm{IR}$ using templates typical of high-redshift star-forming galaxies, hence with higher dust temperatures ($T_\mathrm{d}$). We repeated the calculation using a warmer SED template ($T_\mathrm{d} = 30\, \mathrm{K}$) obtained by scaling the \cite{Magdis2021} SED with a modified blackbody. This yielded even lower maximum L$_\mathrm{IR}$ and more stringent $\mathrm{SFR < 11\, M_\odot\, yr^{-1}}$.

\begin{figure}
    \centering
    \includegraphics[width=0.9\linewidth]{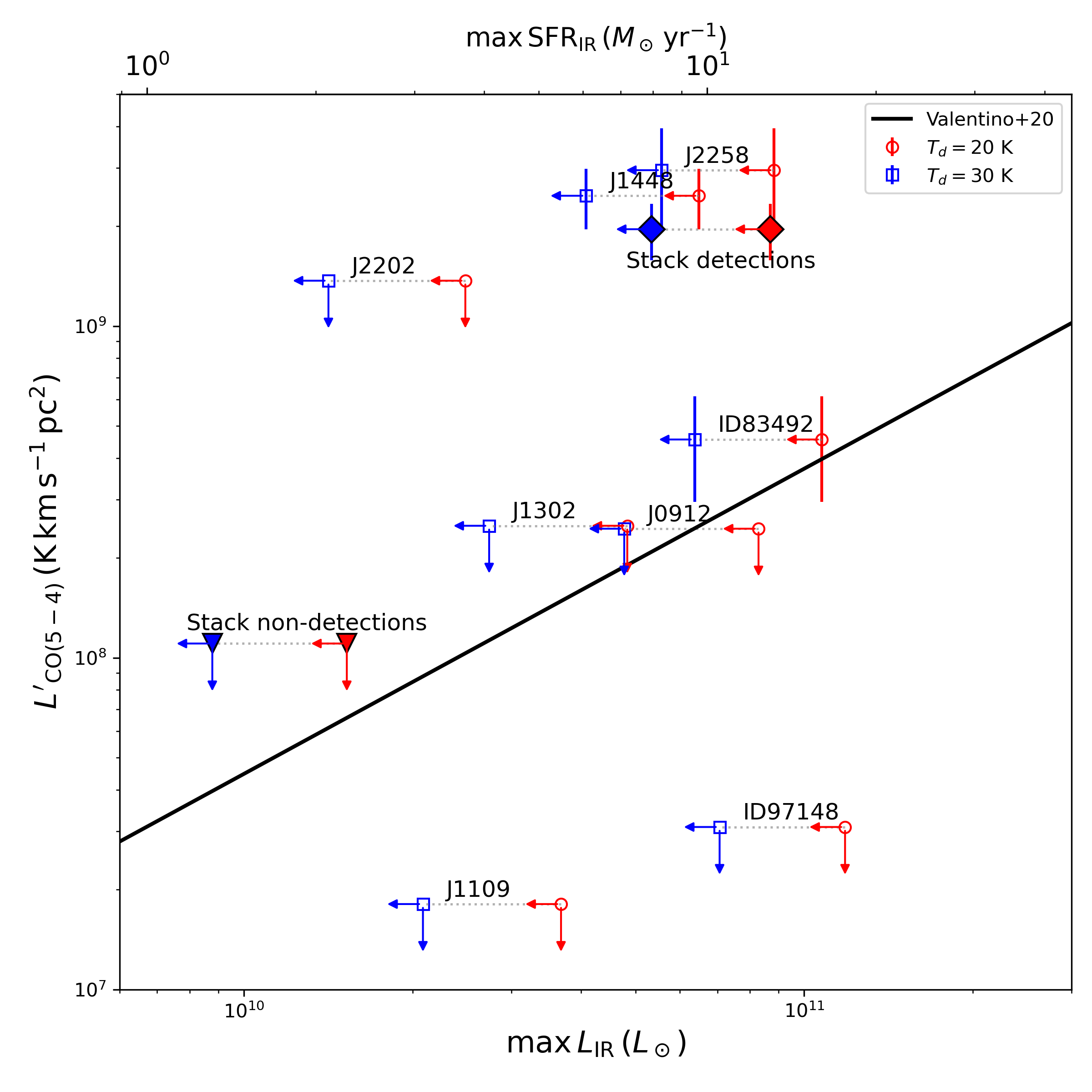}
    \caption{CO(5-4) luminosity of our sample compared with the maximum IR luminosity allowed by the continuum non detections. We computed the maximum L$_\mathrm{IR}$ by considering the average SED template from \cite{Magdis2021} and rescaling it to match the observations. We considered both a template with dust temperature of 20 K (red circles) and 30 K (blue squares, see Sec. \ref{subsec:residual_sf}). We report both individual measurements (empty symbols) and stacking (filled diamonds for detections, down-pointing triangles for non-detections). We plot (black line) the CO(5-4) - L$_\mathrm{IR}$ relation estimated by \cite{Valentino2020b} for star-forming galaxies. The top axis indicates the SFR estimated from the maximum $\mathrm{L_{IR}}$ with the \cite{Kennicutt2021} conversion.}
    \label{fig:maxLIR}
\end{figure}

In Figure \ref{fig:maxLIR} we compare the observed CO(5-4) luminosity with the maximum L$_\mathrm{IR}$ allowed by the dust continuum non-detections. The majority of the measurements do not follow the CO(5-4) - $\mathrm{L_{IR}}$ relation calibrated for star-forming galaxies \citep{Valentino2020b}. For dust temperatures of 20 K, only one detected source and three non-detections are potentially consistent with the relation; for a warmer SED with $T_{\rm d}=30$ K, this number further decreases. To place tighter constraints on the average properties of the two populations, we also derive stacked measurements separately for detections and non-detections. We find that the stack of the detected galaxies is inconsistent with the relation, while the stack of the non-detections remains potentially consistent. Deeper observations and larger samples are needed to assess whether non-detections lie on the relation or significantly deviate from it.

The mismatch observed for most detections, and potentially also for the non-detections, indicates that the CO(5-4) - $L_{\rm IR}$ relation does not apply to the majority of our sample. This result reinforces our conclusion from Sec. \ref{subsec:sled} that the CO(5-4) emission detected in post-SB galaxies is unlikely to be predominantly powered by ongoing star formation. Instead, in at least some systems, the emission is more plausibly driven by alternative excitation mechanisms, such as shocks associated with mergers and/or AGN-related processes. More generally, our results highlight that CO(5-4) - $\mathrm{L_{IR}}$ scaling relations calibrated on star-forming galaxies must be applied with caution when interpreting quiescent or post-starburst systems. This interpretation is consistent with recent findings suggesting that gas and dust in post-starburst and quiescent galaxies are decoupled, exhibiting gas-to-dust ratios that are orders of magnitude higher than in star-forming galaxies \citep{Spilker2025}.
We report all the $L_{\mathrm{IR}}$ and SFR measurements in Table \ref{tab:lir_sfr}.

\section{Discussion}
\label{sec:discussion}

Our post-SB galaxies exhibit an average $R_{52} \sim 0.28$ (from survival analysis) comparable to high-redshift star-forming galaxies. 
However, we find a significantly lower ratio ($R_{52} < 0.10$) when stacking only the galaxies that are not detected in CO(5-4) and do not show signs of interaction, a value twice lower than local star-forming galaxies and 2.5 times lower than high-redshift ones. This suggests a reduced fraction of dense and/or warm molecular gas relative to cold and diffuse gas in these sources, consistent with previous studies of local early-type \citep{Crocker2012, Bayet2013} and post-SB galaxies \citep{French2023}.

As highlighted by \cite{French2023}, the excitation of CO-traced molecular gas can help distinguish between different mechanisms responsible for the low SFE observed in post-SBs. Low SFE is thought to arise from a deficit of dense molecular gas; however, the excitation of the remaining gas depends on its kinetic temperature and density, and thus provides insight into the underlying physical processes. Elevated kinetic temperatures, potentially driven by turbulence, shocks, or AGN activity, would result in high excitation, whereas low-density gas with modest temperatures would produce low excitation levels similar to those observed in early-type galaxies.

The average CO excitation of the post-starburst galaxies in our sample favors this latter scenario, pointing to molecular gas that is predominantly diffuse, cold, and inefficient at forming stars. This interpretation is consistent with the low-excitation CO SLEDs observed in local post-SBs by \cite{French2023}, who constrained transitions up to CO(3–2), and supports a picture in which quenching is maintained by a deficit of dense star-forming gas rather than by widespread heating alone. Such a low dense-gas fraction suggests that gas stabilization processes, such as morphological quenching \citep{Martig2009, Johansson2009} or enhanced turbulence that prevents the collapse of gas into giant molecular clouds \citep{Smercina2021}, play a dominant role in suppressing star formation in these systems.

For the three post-SBs in our sample that are currently undergoing a merger, however, the physical conditions of the molecular gas may be different. Although their $R_{52}$ values are comparable to those of main-sequence galaxies, this similarity does not necessarily imply a common excitation mechanism. In these systems, the elevated high-$J$ CO emission may instead arise from enhanced heating, potentially driven by merger-induced shocks and/or AGN activity.
Galaxy mergers are known to strongly perturb the interstellar medium, generating shocks and turbulence that can modify the gas density and temperature distribution. In particular, compressive turbulence during mergers can increase the fraction of dense gas \citep[e.g.][]{Barnes1991, Jog1992, Renaud2014, Renaud2022}, especially during the first pericenter passages, in contrast to the predominantly solenoidal turbulence inferred in local post-SBs by \cite{Smercina2021} and \cite{French2023} that suppresses star formation while maintaining low CO excitation. High-resolution simulations by \cite{Bournaud2015} show that major mergers exhibit significantly more excited CO SLEDs than local spirals, with CO(5–4)/CO(1–0) intensity ratios on average a factor of $\sim$8 higher in starbursting mergers. This enhanced excitation is attributed to dense gas created by large-scale inflows and merger-driven turbulence on small scales. Observational studies of local mergers similarly report elevated CO excitation associated with warm and/or dense molecular gas, potentially linked to enhanced turbulence and cosmic-ray heating \citep{Papadopoulos2012, Liu2021}. We also notice that for the two sources in our sample (J1448 and J2258) with the highest excitation, the CO(5-4) emission has broader linewidth than the CO(2-1), further indicating that the kinematics of the warm gas phase might be perturbed by merger-driven turbulence, inflows, outflows, and/or shocks.

In at least one of the merging post-SBs in our sample, we also find evidence for AGN activity. Both X-ray irradiation from the AGN and shocks associated with outflows have been shown to enhance CO excitation \citep{Vallini2019}, and may contribute to the steeply rising SLED of J1448. High-resolution morphological and kinematic observations of the CO-emitting gas will be essential to determine the spatial origin of the dense and/or warm molecular component in these systems and to disentangle the respective roles of turbulence, mergers, and AGN feedback in maintaining quenching.

Interestingly, the companion galaxies of ID83492 show $R_{52}$ values 0.5 dex lower than those of high-redshift starbursts with comparable sSFRs. This likely reflects a mismatch in timescales: while sSFR traces star formation averaged over tens of Myr, CO(5–4) probes the instantaneous reservoir of dense and warm gas. In a merger-driven starburst, the dense gas phase can be rapidly consumed or disrupted \citep[e.g.,][]{Petersson2023}, leading to low $R_{52}$ even while the galaxy still appears star-forming. Future observations of $\mathrm{H\alpha}$ or $\mathrm{Pa\alpha}$ lines will help constrain the instantaneous SFR and assess whether a timescale mismatch indeed underlies the apparent tension between the $R_{52}$ and sSFR in these systems. Additionally, observations of multiple CO transitions will be key to better constrain the SLEDs of these galaxies.

 We note, however, that our analysis is limited to post-starbursts with existing low-$J$ CO detections, which may represent the gas-rich tail of the population. The processes responsible for shutting down star formation and those that subsequently maintain quiescence may therefore differ in gas-poor passive galaxies. Spatially resolved observations would provide stronger constraints by mapping the extent of both high-$J$ and low-$J$ CO emission and identifying potential spatial variations in excitation across the galaxies. In addition, higher signal-to-noise data would enable dynamical studies to assess whether the molecular gas is gravitationally stable and to search for evidence of outflows.

None of the galaxies in our sample are detected in the dust continuum, despite the fact that, based on their CO(5-4) detections and the CO(5-4) - L$_\mathrm{IR}$ relation from \citet{Valentino2020b}, we would have expected at least some detections. This result suggests that the CO(5-4) - L$_\mathrm{IR}$ relation established for star-forming galaxies may not hold for quiescent systems, and that molecular gas and dust could be decoupled in passive galaxies. This interpretation is consistent with recent studies reporting unexpectedly weak dust emission and unusually high gas-to-dust ratios ($\delta_\mathrm{GDR} \sim 300$ - $1200$; \citealt{Spilker2025}). These findings imply that dust continuum emission is an unreliable tracer of molecular gas in quiescent galaxies and highlight the need for alternative cold-gas diagnostics (e.g., CO). Future mid-IR observations of post-SB and quiescent systems will provide further insight into their SEDs, constrain their dust properties, and potentially reveal residual obscured star formation or AGN activity.

\section{Summary and conclusions}
\label{sec:conclusions}

We investigated the molecular gas excitation of eight post-starburst galaxies at $z \sim 0.6 - 1.3$ by comparing CO(5-4) emission with archival CO(2-1), CO(3-2), and CO(4-3) data. To our knowledge, this is the first time that high-$J$ CO transitions are probed for quiescent galaxies at these redshifts. All galaxies in our sample are detected in low-$J$ CO transitions and have molecular gas fractions up to $\sim 30\%$ \citep{Bezanson2021, Zanella2023}. Our main findings are as follows:

\vspace{-0.3cm}
\begin{itemize}
\item Three out of the eight galaxies are detected in CO(5-4). All three show signs of ongoing mergers or interactions, either in the form of tidal tails (J1448 and J2258; \citealt{Spilker2022, Donofrio2025}) or via two nearby star-forming companion galaxies (ID83492; \citealt{Zanella2023}).

\item We used the $R_{52} = L'_\mathrm{CO(5-4)}/L'_\mathrm{CO(2-1)}$ ratio as a proxy for molecular gas excitation. On average, our post-starburst galaxies have $R_{52} = 0.28$ (from survival analysis), comparable to high-redshift main-sequence galaxies. However, when considering only the non detections (coinciding also with sources that do not show signs of interaction), we estimate a significantly lower $R_{52}< 0.10$ (from stacking), 2 times lower than local star-forming galaxies  and $>2.5$ times lower than high-redshift main-sequence and starburst galaxies. In contrast, the three merging post-starbursts exhibit higher $R_{52} \sim 0.49$, more similar to high-redshift starbursts. Interestingly, the two star-forming companions of ID83492, which have sSFRs comparable to starburst galaxies, show $R_{52}$ values $\sim 0.5$ dex lower than typical main-sequence sources.

\item The CO SLEDs of our post-SBs are diverse, with an average peak at $J = 3$, similar to the Milky Way. The merging galaxies, however, have SLEDs peaking at higher $J$, more akin to high-redshift main-sequence galaxies, with J1448 being the most extreme case, likely requiring additional mechanisms (e.g., shocks or AGN activity) to explain its elevated excitation.

\item Residual obscured star formation cannot fully account for the CO(5-4) emission in the detected galaxies, because assuming the CO(5-4) - $L_\mathrm{IR}$ relation for star-forming galaxies \citep{Valentino2020b} and the typical SED of quiescent galaxies \citep{Magdis2021}, they would have been detected in the dust continuum. Additionally, the fact that our post-starbursts do not agree with the CO(5-4) - L$_\mathrm{IR}$ relation and that none of them is detected in dust continuum indicates that molecular gas and dust in quiescent galaxies might be decoupled, as recently suggested by \cite{Spilker2025}.

\item Our results indicate that, for the majority of our post-SBs, the molecular gas is predominantly diffuse and cold, leading to low CO excitation consistent with that observed in local post-SBs \citep{French2023}. This suggests that the suppression of star formation is primarily maintained by a low fraction of dense molecular gas, favoring quenching mechanisms such as gas stabilization or enhanced turbulence that prevent the gas from fragmenting and collapsing, rather than by widespread heating of the molecular medium. For the merging post-starbursts, however, the higher CO excitation indicates elevated kinetic temperatures, likely induced by shocks and/or AGN activity, which act to suppress star formation.
\end{itemize}

Larger samples of high-redshift post-starbursts with multi-transition CO observations will be essential to confirm and expand upon these results. Recently, \cite{Setton2025} published a sample of 50 post-SBs with CO(2-1) data that could in principle be followed-up with observations of higher $J$ CO transitions. In particular, homogeneously observing multiple CO transitions across large samples will allow statistical constraints on the SLED and help determine whether elevated excitation is common in merging post-SBs. In addition, further investigating the SLED of passive galaxies might shed light on the $\alpha_\mathrm{CO}$  factor that needs to be used to convert low-$J$ CO luminosities into molecular gas masses. Current works (e.g., \citealt{French2015, French2021, Suess2017, Bezanson2021, Zanella2023, Setton2025}) typically adopt the Milky-Way conversion factor. However, $\alpha_\mathrm{CO}$ might depend on the SLED of the galaxy population and could therefore be different in quiescent galaxies. Higher-resolution observations will further enable measurements of the CO-emitting region’s size, morphology, and dynamics, and help deblend close galaxy pairs, clarifying whether mergers are ubiquitous in the post-SB population and their role in suppressing star formation. Finally, assessing the presence of AGN and potential molecular outflows will be critical to understanding the contribution of AGN feedback and shocks to quenching.

\begin{acknowledgements}
We thank the referee for the insightful report.
AZ thanks F. Vito for useful discussions about the X-ray detection and AGN activity of our galaxies. SB is supported by ERC grant 101076080. FV acknowledges support from the Independent Research Fund Denmark under grant 3120-00043B and the Danish National Research Foundation under grant DNRF140. This paper makes use of the following ALMA data: 2016.1.01126.S; 2017.1.01109.S; 2018.1.01264.S; 2018.1.01240.S; 2019.1.00221.S; 2019.1.00900.S; 2024.1.00061.S. 
ALMA is a partnership of ESO (representing its member states), NSF (USA), and NINS (Japan), together with NRC (Canada), MOST and ASIAA (Taiwan), and KASI (Republic of Korea), in cooperation with the Republic of Chile. The Joint ALMA Observatory is operated by ESO, AUI/NRAO and NAOJ.
\end{acknowledgements}

\section*{Data Availability}

The data used in this study are available from telescope archives. Software and derived data generated for this research can be made available upon reasonable request to the corresponding author.

\bibliographystyle{aa}
\bibliography{bibliography} 

\begin{appendix}

\section{Continuum detections}
\label{app:continuum_det}

We searched for dust continuum detections nearby our targets and found several targets. We detected the following sources:

\begin{itemize}
\item four galaxies with a projected distance of $6.0\arcsec - 15.3\arcsec \sim 50.5 - 129.6\, \mathrm{kpc}$ from ID83492. All of them are detected in \textit{HST}/F160W optical imaging. The dust continuum detection of two of them, the brightest, was already reported by \cite{Zanella2023}. They are also detected in CO(2-1) and CO(3-2) at a similar redshift as our main target. The remaining two do not show any emission lines in ALMA data. They have photometric redshift from 3D HST \cite{Skelton2014} $\mathrm{z_{phot} = 1.9808}$ and $\mathrm{z_{phot} = 2.0229}$, hence they are likely at higher redshift than our target.
\item one galaxy at $9.7\arcsec \sim 83.4\, \mathrm{kpc}$ from ID97148 ($5.4\sigma$). This source is detected in optical VIRCAM $H$ band data and it has several photometric redshift estimates from the literature in the range $\mathrm{z_{phot} = 2.0 - 2.4}$ \citep{Mehta2018, Hatfield2022}, indicating that this is likely a higher redshift source than our target. 
\item one galaxy at $5.4\arcsec \sim 36.8\, \mathrm{kpc}$ from J1302 ($5.4\sigma$). The available photometric redshift estimate from the DESI survey $\mathrm{z_{phot} = 0.78}$ \citep{Duncan2022} indicates that this is likely a lower redshift galaxy than J1302.
\end{itemize}

We report in Table \ref{tab:continuum} their coordinates and fluxes.

\begin{table}[h!]
    \centering
    \caption{Coordinates and continuum flux of the sources serendipitously detected nearby our targets.}
    \begin{tabular}{c c c c}
    \toprule
    \midrule
     ID    & RA    & Dec    & F$_\mathrm{cont}$  \\
           & (deg) & (deg)  & ($\mu$Jy)              \\
    \midrule
    ID83492-cont1 & 34.548128 & -5.151115 & $67.4 \pm 1.0$ \\
    ID83492-cont2 & 34.549892 & -5.144405 & $72.0 \pm 14.7$ \\
    ID97148-cont1 & 34.332948 & -5.081456 & $71.0 \pm 13.2$ \\
    J1302-cont1   & 195.704078 & 10.718890 & $87.7 \pm 16.2$ \\
    \bottomrule
    \end{tabular}
    \label{tab:continuum}
\end{table}

\section{Molecular gas excitation}
\label{app:excitation}

\begin{figure}[th!]
    \centering
    \includegraphics[width=\linewidth]{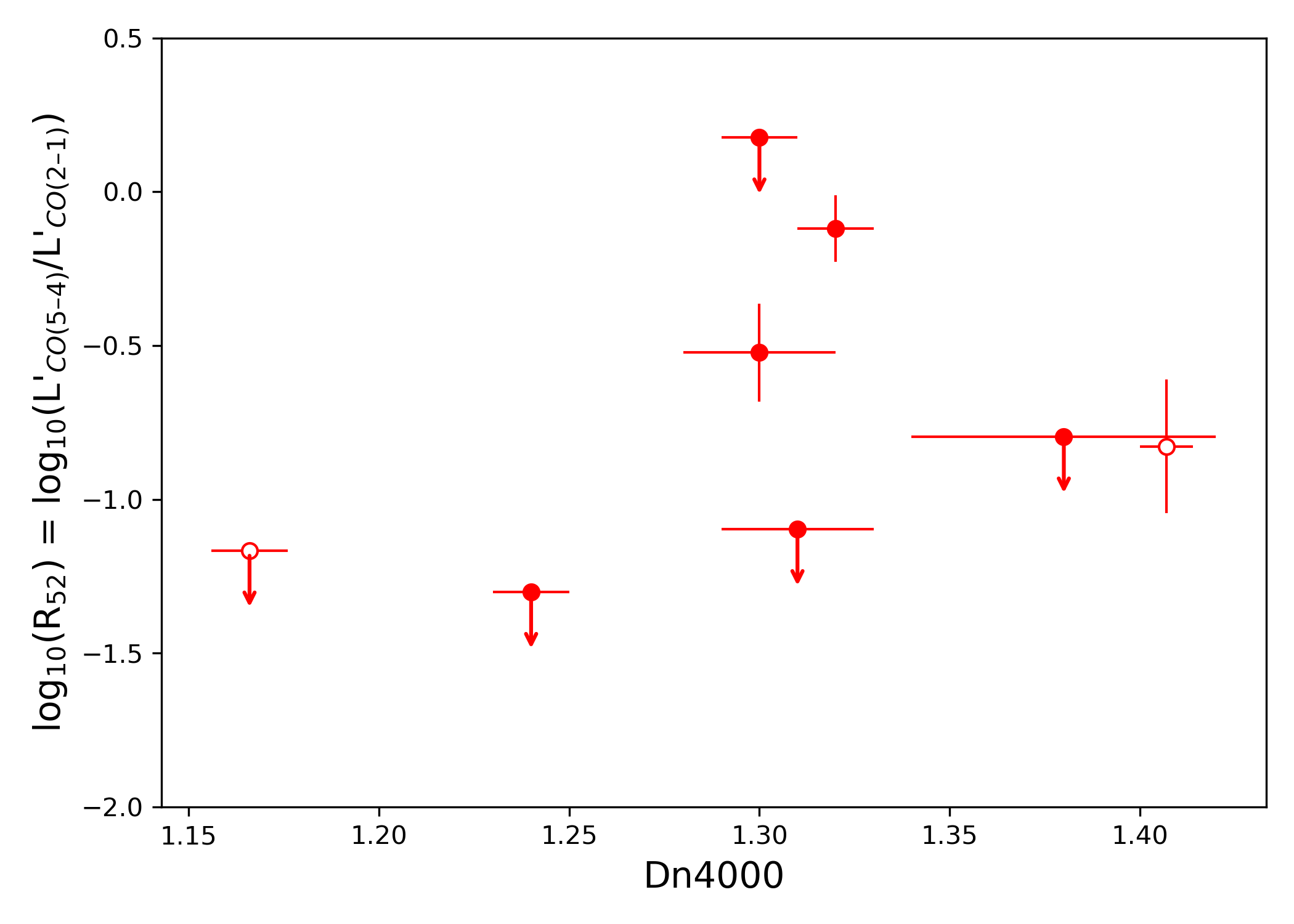}
    \includegraphics[width=\linewidth]{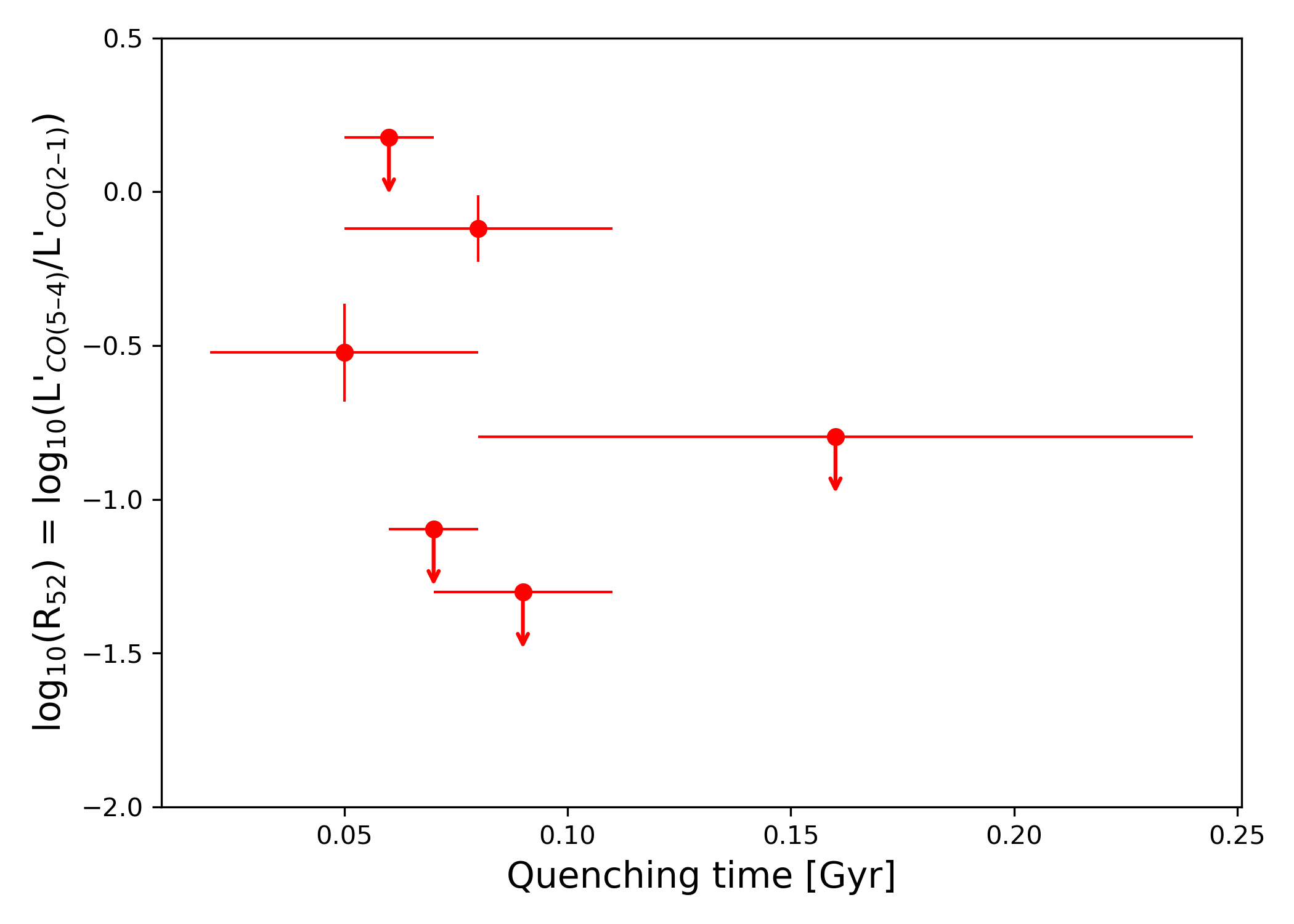}
    \includegraphics[width=\linewidth]{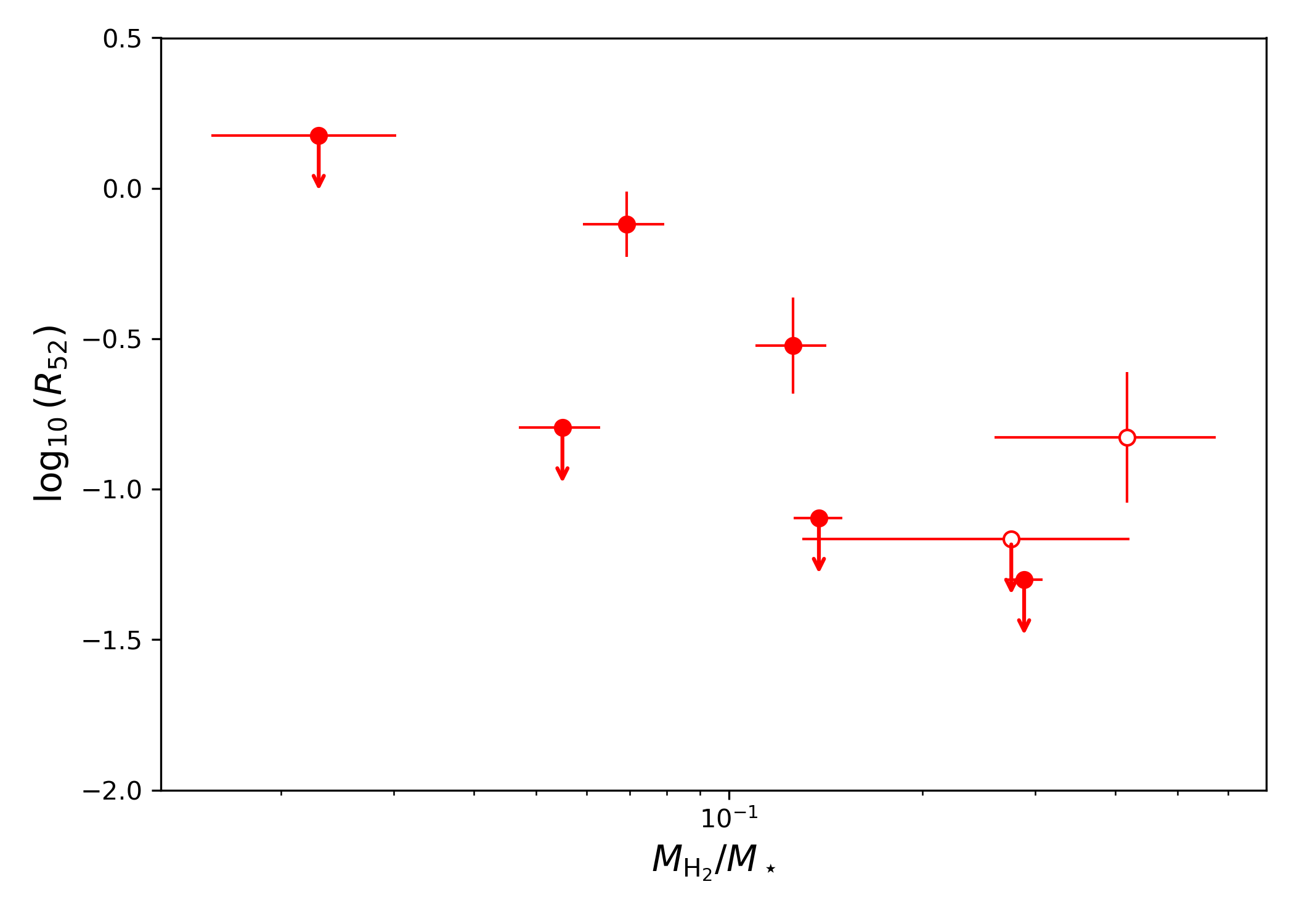}
    \caption{Molecular gas excitation traced by the CO(5-4)/CO(2-1) brightness temperature ratio ($R_{52}$) as a function of galaxy properties. \textit{Top left}: $R_{52}$ versus the 4000\AA\,  break. \textit{Top right}: $R_{52}$ versus the time since quenching. \textit{Bottom}: $R_{52}$ versus the molecular gas fraction.}
    \label{fig:co_ratio_properties}
\end{figure}

We investigated possible correlations between the CO(5–4)/CO(2–1) luminosity ratio and several galaxy properties, namely the 4000 \AA\, break strength (Dn4000), the quenching timescale ($t_\mathrm{q}$), and the molecular gas fraction ($M_{\mathrm{H2}}/M_\star$), as shown in Fig.~\ref{fig:co_ratio_properties}.
To quantify any trends, we computed both the Pearson and Spearman correlation coefficients. 
We found no statistically significant correlations for Dn4000, $r_\mathrm{P}=0.31$ ($p=0.46$) and for $t_\mathrm{q}$, $r_\mathrm{P}=-0.32$ ($p=0.54$). The corresponding Spearman coefficients are similarly low ($|r_\mathrm{S}|<0.49$) with $p>0.32$. The only marginally significant correlation iw with $M_{\mathrm{H2}}/M_\star$, which gives Pearson coefficients $r_\mathrm{P}=-0.64$ ($p=0.0.9$) and Spearman coefficients $r_\mathrm{S} = -0.79$) with $p = 0.02$.  These results indicate that, within the current uncertainties and limited sample size, there is no compelling evidence for a correlation between the CO excitation (as traced by $R_{52}$) and stellar population age or quenching timescale, and a marginal correlation with the gas content. The latter is to be explored with future datasets.

\section{Residual obscured star formation}
\label{app:obscured_sfr}

We investigated whether our sample galaxies had residual star formation by converting the CO(5-4) luminosity into total IR luminosity following the relation of \cite{Valentino2020b}. We then converted the L$_\mathrm{IR}$ into SFR by using the \cite{Kennicutt2021} relation (see Section \ref{subsec:residual_sf} for details). We also computed the maximum L$_\mathrm{IR}$ consistent with the dust continuum non detection considering the average SED template for quiescent galaxies from \cite{Magdis2021} both with a dust temperature of 20 K and 30 K. In Table \ref{tab:lir_sfr} we report the L$_\mathrm{IR}$ and SFR obtained with the different methods.

\begin{table*}[]
    \centering
    \caption{IR luminosities and star formation rates.} 
    \begin{tabular}{c c c c c c c}
    \toprule
    \midrule
    ID     & $\mathrm{L_{IR,V20}}$ & $\mathrm{SFR_{V20}}$ & $\mathrm{L_{IR,20K}}$ & $\mathrm{SFR_{20K}}$ & $\mathrm{L_{IR,30K}}$ & $\mathrm{SFR_{30K}}$ \\
           & ($\mathrm{10^{10}L_\odot}$) & (M$_\odot$ yr$^{-1}$) & ($\mathrm{10^{10}L_\odot}$) & (M$_\odot$ yr$^{-1}$) & ($\mathrm{10^{10}L_\odot}$) & (M$_\odot$ yr$^{-1}$) \\
    \midrule
    J1448   & $78.4 \pm 17.6$ & $116.7 \pm 26.3$ & $< 6.5$ & $< 9.7$ & $< 4.1$ & $< 6.1$\\
    J2258   & $95.1 \pm 35.0$ & $141.6 \pm 52.2$ & $< 8.8$ & $< 13.2$ & $< 5.6$ & $< 8.3$\\
    ID83492 & $12.5 \pm 4.8$ & $18.6 \pm 7.1$ & $< 10.8$ & $< 16.0$ & $< 6.4$ & $< 9.5$\\
    J1109   & $< 2.0$ & $< 2.9$ & $< 3.7$ & $< 5.5$ & $< 2.1$ & $< 3.1$\\
    J1302   & $< 3.9$ & $< 5.9$ & $< 4.8$ & $< 7.2$ & $< 2.7$ & $< 4.1$\\
    J0912   & $< 3.7$ & $< 5.5$ & $< 8.3$ & $< 12.3$ & $< 4.8$ & $< 7.1$\\
    J2202   & $< 23.9$ & $< 35.6$ & $< 2.5$ & $< 3.7$ & $< 1.4$ & $< 2.1$\\
    ID97148 & $< 0.8$ & $< 1.2$ & $< 11.8$ & $< 17.6$ & $< 7.1$ & $< 10.5$\\
    \midrule
    Mean mergers & $60.9 \pm 12.8$ & $90.7 \pm 19.0$ & $< 3.1$ & $ < 4.6$ & $< 1.8$ & $< 2.7$ \\
    Stack isolated & $< 2.6$ & $< 3.9$ & $< 1.5$ & $< 2.3$ & $< 0.9$ & $< 1.3$ \\
    \bottomrule
    \end{tabular}
    \label{tab:lir_sfr}
    \tablefoot{IR luminosity and star formation rate have been obtained \textit{(i)} converting the CO(5-4) luminosity into L$_\mathrm{IR}$ using the \cite{Valentino2020b} relation ($\mathrm{L_{IR,V20}}$, $\mathrm{SFR_{V20}}$); \textit{(ii)} considering the observed continuum upper limits and the average SED template for quiescent galaxies from \cite{Magdis2021}. We considered both a template dust temperature of 20 K ($\mathrm{L_{IR,20K}}$, $\mathrm{SFR_{20K}}$) and 30 K ($\mathrm{L_{IR,30K}}$, $\mathrm{SFR_{30K}}$).}
\end{table*}

\end{appendix}

\end{document}